\documentclass[dvipsnames, journal]{vgtc}              % review (journal style)
\usepackage{xurl}

\onlineid{2151}

\newcommand\hide[1]{}

\newcommand\revision[1]{{#1}}

\newcommand\jf[1]{\textcolor{red}{#1}}
\newcommand\scq[1]{\textcolor{olive}{#1}}
\newcommand\ew[1]{\textcolor{cyan}{}}
\newcommand\joao[1]{\textcolor{purple}{}}

\renewcommand\jf[1]{}
\renewcommand\scq[1]{}
\renewcommand\ew[1]{}
\renewcommand\joao[1]{}

\renewcommand\paragraph[1]{\noindent \textbf{#1}}

\newcommand{\name}{\emph{UrbanTrace}\xspace}
\newcommand{\compCanvas}{\emph{Spatial Integration Canvas}\xspace}
\newcommand{\compAtlas}{\emph{Atlas Profiler}\xspace}
\newcommand{\compDiscoveryAgent}{\emph{Semantic Data Discovery Agent}\xspace}
\newcommand{\compOperationCopilot}{\emph{Semantic-Aware Operator Copilot}\xspace}
\newcommand{\compSynthesisAgent}{\emph{Synthesis
Agent}\xspace}

\newcommand{\autoDdg}{\emph{AutoDDG}\xspace}

\usepackage{xcolor}

\newcommand{\chip}[3]{%
  {\setlength{\fboxsep}{1.2pt}%
  \colorbox{#2}{\textcolor{#3}{\scriptsize\textsf{\textbf{#1}}}}}%
}

\newcommand{\cdatadiscovery}{%
  \texorpdfstring
    {\chip{C1}{orange!18}{orange!70!black}}
    {C1}%
}

\newcommand{\cspatialunit}{%
  \texorpdfstring
    {\chip{C2}{blue!14}{blue!65!black}}
    {C2}%
}

\newcommand{\cmeasurement}{%
  \texorpdfstring
    {\chip{C3}{purple!14}{purple!65!black}}
    {C3}%
}

\vgtccategory{Research}

\title{\name: LLM-Assisted Discovery and Semantics-Aware Integration of Spatial Data}

\author{%
\authororcid{Sonia Castelo}{0000-0001-6881-3006}, 
\authororcid{Eden Wu}{0009-0003-6455-917X}, 
\authororcid{Jo\~ao Rulff}{0000-0003-3341-7059},
\authororcid{Harish Doraiswamy}{0000-0003-2995-250X}, 
\authororcid{Juliana Freire}{0000-0003-3915-7075}, 
\authororcid{Cl\'audio Silva}{0000-0003-2452-2295} 
}
\authorfooter{
\item Authors are with the New York University (NYU) and Microsoft Research. E-mails: \{s.castelo, eden.wu, jlrulff, juliana.freire, csilva\}@nyu.edu. \{harish.doraiswamy\}@microsoft.com
}

\abstract{%
Urban decision-making requires integrating heterogeneous spatial data. While current GIS tools handle geometric computation efficiently, they lack the semantic reasoning to guide complex workflows. Analysts manually \revision{manage} data discovery, spatial boundaries, and measurement semantics,
risking \hide{silent} aggregation errors. We present \name, a visual analytics system that transforms manual spatial data-wrangling into a transparent, node-based collaborative workflow with context-aware AI agents. Using an offline profiler to extract semantic and geometric metadata, \name grounds LLMs in real-world data distributions. This enables specialized agents to retrieve datasets based on high-level goals and automatically enforce valid spatial aggregations. To make harmonization explicit, \revision{three interactive views: an Integration Provenance Graph, Multivariate Priority Map, and Spatial Delta Map,} allow users to explore how conclusions shift across spatial configurations. We evaluate \name \revision{ on 28 urban scenarios spanning 112 datasets.}
Quantitative ablations show our profiling significantly outperforms baseline LLMs in data discovery, achieving 100\% semantic and 87\% geometric validity in spatial mapping. \revision{Through} real-world case studies and expert interviews, we demonstrate that \name turns spatial aggregation sensitivity from a methodological burden into an exploratory visual asset.
}


\teaser{
  \centering
    \includegraphics[width=\linewidth, alt={A view of a city with buildings peeking out of the clouds.}]{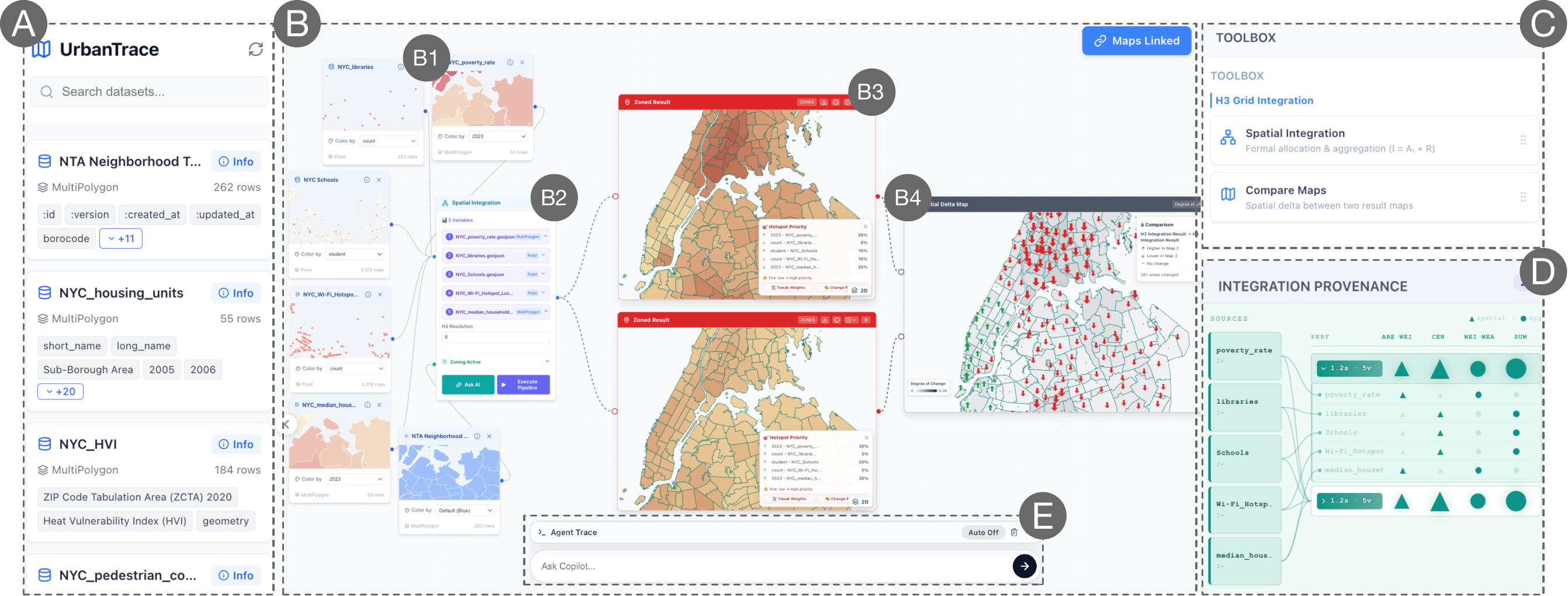}
  \caption{%
  	   \name applied to a real-world analysis identifying NYC neighborhoods in need of after-school programs. The workflow begins with the (A) Dataset Selector and is supported by the (E) \revision{\compDiscoveryAgent}, which recommends datasets for the analysis. The selected datasets are displayed on the (B) Main Canvas as (B1) Dataset Nodes. These are connected to (B2) Integration Nodes, where \revision{ a \compOperationCopilot} recommends valid spatial transformations. The provenance of the transformations is visualized in the (D) \revision{Integration Provenance Graph} (rectangles: data, triangles: mapping operators, circles: aggregations) -- hovering over a transformation reveals the AI-generated rationale. The \revision{\compSynthesisAgent} then generates a (B3) Result Map Node that displays \revision{a Multivariate Priority Map}, with hotspots indicating areas of need. Using the (C) Spatial Analysis Panel, analysts can iteratively test counterfactuals by modifying weights or operators, and compare outcomes via the (B4) Spatial Delta Map, which highlights the differences between transformations, i.e., emerging (green) and diminishing (red) priority hotspots.
  }
  \label{fig:teaser}
}

\graphicspath{{figs/}{figures/}{pictures/}{images/}{./}} % where to search for the images

\usepackage{tabu}                      % only used for the table example
\usepackage{booktabs}                  % only used for the table example
\usepackage{lipsum}                    % used to generate placeholder text
\usepackage{mwe}                       % used to generate placeholder figures
\usepackage{ccicons}                   % package to be able to use icons from creative commons

\usepackage{mathptmx}                  % use matching math font
\usepackage{amsmath}                   % for \text in math mode
\usepackage{amsfonts}
\usepackage{algorithm}
\usepackage{algorithmic}

\begin{document}

\firstsection{Introduction}
\label{sec:introduction}

\maketitle

Urban decision-making increasingly depends on integrating heterogeneous spatial datasets -- combining census surveys, administrative records, and infrastructure maps to understand where need is concentrated and how resources should be allocated. 
%
% \jf{this sentence sounds weird: but the reasoning required to use them correctly does not --pls check my rewording} \scq{done.}
\revision{There are analytical tools for these tasks}~\cite{arcgis, sedona_geospark, geopandas}, %, Rey2010PySAL}, 
%the reasoning required to use them correctly does not: 
\revision{but users struggle with identifying datasets that are relevant to a given question,
determining how to join datasets with incompatible zoning systems, and \hide{how to} properly aggregate different variables.
%which datasets are relevant to a given question, how they should be joined across incompatible zoning systems, and whether the variables they contain can be combined under a given aggregation operation. 
These decisions are typically made manually, in an ad-hoc fashion.} We illustrate these \revision{challenges} through a concrete example.
%to design and implement evidence-based strategies. Whether identifying areas for infrastructure investment or simulating the impact of climate events, analysts must reconcile disparate data sources to build a unified spatial dataset. However, the path from a high-level analytical question to a harmonized dataset is fraught with manual data-wrangling hurdles. Existing Geographic Information Systems (GIS) provide the computational tools for these tasks but fail to automate the reasoning required to navigate them, leaving analysts in a high-friction process of manual discovery and ad-hoc integration.

%When analyzing spatial data, we generate summaries at geographic levels — census tracts, ZIP codes, neighborhoods, and boroughs. These zones are determined by administrative needs and data-collection logistics. While they are arbitrary with respect to the analysis question, using different zones -- with different sizes or shapes --  yields different answers. This is the infamous MAUP problem~\cite{WONG2009MAUP}. To do any joint analysis, one must move multiple datasets into common set of zones. Since there is no single correct answer to the question "what zones should I use?"  it is important for analysts to explore different choices and their implications.

Suppose an analyst is tasked with implementing New York City's Universal After-School initiative \cite{nyc_after_school_2025}. To allocate resources effectively, they must answer a specific operational question: \emph{which neighborhoods should be prioritized for additional after-school programs?} Answering this question immediately surfaces three interrelated challenges.

\paragraph{\cdatadiscovery~Challenge 1: Data Discovery.}
The first question \revision{to} be answered is which datasets are needed for the analysis.
%Answering this prioritization question requires 
An expert would look for data on student demand, existing digital and educational infrastructure, and socio-economic risk. But these are not stored in a single dataset. School enrollment, library locations, and public Wi-Fi spots are maintained as discrete point data across different municipal portals, such as Capital Planning NYC and NYC Open Data~\cite{capitalplanning,nycopendata}. Meanwhile, socio-economic indicators \revision{such as} poverty and income rates are published by the Furman Center~\cite{furmancenter2024} \revision{and} aggregated \revision{into} multipolygons at the Community District level. Each dataset was collected for different administrative purposes and different datasets use different spatial representations. Given that dataset descriptions are often terse~\cite{zhang2025autoddg}, it is difficult to design keyword queries that retrieve the desired datasets. 
% \jf{can we give an example as to why simple keyword search would not work? for example, that the title or description does not contain an intuitive keyword?} \scq{done}
For example, an analyst searching for ``internet access'' may miss relevant data if it appears under administrative titles like ‘LinkNYC Kiosk Locations’ without intuitive keywords. There is no systematic guidance on which datasets are relevant to a given question or how they relate to one another. Analysts must discover, evaluate, and assemble them through a combination of domain expertise and manual search, a process that remains time-consuming, error-prone, and largely invisible in the final analysis, despite advances in data cataloging and profiling~\cite{feuer2023archetype, castelo2021auctus, fernandez2018aurum,deng2017thedatacivilizer}. 

\paragraph{\cspatialunit~Challenge 2: Selection of Spatial Unit for Joint Analysis.} Once the relevant datasets are selected, combining point-level infrastructure (e.g., schools, libraries) with regional indicators (e.g., poverty rates) requires choosing a common spatial unit. This choice is never neutral, as different boundaries yield 
%fundamentally 
different answers, the Modifiable Areal Unit Problem (MAUP)~\cite{WONG2009MAUP}.
For example, aggregating library access to a large Community District might mask localized deficits, while granular census tracts might artificially isolate a school from a library just across the street.
% Consider a localized deficit of public libraries near a cluster of schools: aggregating to a large Community District might smooth this deficit away using a surplus from the opposite side of the district, whereas analyzing at the granular census-tract level might artificially classify a school as critically under-resourced simply because the nearest library falls just across the street into an adjacent tract.
Because administrative zones (e.g., census tracts, Council Districts) 
% , whether census tracts, Neighborhood Tabulation Areas (NTAs), or City Council Districts,
were drawn for purposes unrelated to after-school planning, the answer to the operational question changes with the zoning system, not the underlying reality. Consequently, no single ``correct'' spatial unit exists. Analysts must instead explore how conclusions shift across different aggregations~\cite{goodchild1980areal, flowerdew1991areal}, treating boundary sensitivity as a valuable analytical signal rather than statistical noise.

\paragraph{\cmeasurement~Challenge 3: Measurement Semantics.}
% Not all variables behave the same way under spatial aggregation, 
\revision{Variables behave differently under spatial aggregation,}
and applying the wrong operation produces %silently
incorrect results~\cite{STASCH2014Meaningfulaggregation}. 
For instance, student counts are \emph{extensive} variables that sum correctly across merged neighborhoods. Conversely, poverty rates are \emph{intensive} variables~\cite{pebesma2025spatial, Scheider2019extensiveintensivepropertie} requiring population-weighted averages; 
% summing them is a critical mathematical error that artificially inflates 
\revision{summing them artificially inflates} 
vulnerability in larger districts. 
%Consider the total number of students (School Student Count) versus the percentage of families living in poverty (Poverty Rate). The student count is an \emph{extensive} variable that correctly sums when neighborhoods are merged. Conversely, the poverty rate is an \emph{intensive} variable~\cite{pebesma2025spatial, Scheider2019extensiveintensivepropertie} requiring a population-weighted average; simply summing poverty rates across zones is a critical mathematical error that artificially inflates vulnerability in larger districts. 
This semantic distinction \revision{governs} every step of spatial harmonization, yet most standard GIS tools lack inherent \revision{\emph{semantic awareness} and treat} all numerical columns identically by default. The analyst has no way of knowing which is which unless the measurement type of each variable is explicitly encoded and enforced. Consequently, these three challenges are deeply intertwined: safely exploring spatial boundaries (\cspatialunit) depends on applying the correct semantic aggregation operators (\cmeasurement), 
and both depend on successfully discovering the relevant datasets (\cdatadiscovery).
% and both depend on having first discovered and correctly understood the relevant datasets (Challenge 1).
%
Existing GIS tools~\cite{arcgis, geopandas, Rey2010PySAL}
% (e.g., ArcGIS~\cite{arcgis}, GeoPandas~\cite{geopandas}, PySAL~\cite{Rey2010PySAL}),
%whether commercial platforms like ArcGIS~\cite{arcgis} or open-source libraries like GeoPandas~\cite{geopandas}, and PySAL \cite{Rey2010PySAL},
address the geometric and computational aspects of these tasks but provide no support for the reasoning layer: they cannot identify relevant datasets, infer measurement semantics, or flag when an aggregation operation is semantically invalid.

\paragraph{Our Approach: \name.} This paper presents a framework that makes all three challenges explicit and tractable through \name, a %mixed-initiative 
visual analytics system in which analysts and AI agents collaborate. By leveraging a large language model (LLM) to bridge the gap between high-level analytical goals and harmonized spatial data, \name transforms rigid, ``black-box'' scripts into a transparent workflow. This environment enables analysts to interactively navigate the complexities of data discovery, spatial mapping, and aggregation operators, and to explore how different choices affect outcomes.

A key requirement for data discovery and integration is \emph{data understanding}.
As an offline process, \name applies the \compAtlas to extract semantic and geometric metadata to ground subsequent AI reasoning in statistical and geographic realities.
The derived metadata enable the \compDiscoveryAgent to retrieve datasets based on their analytical relevance to high-level goals such as \emph{``identify neighborhoods underserved by after-school technology programs''}, rather than simple keyword matching. 
Once datasets are selected for analysis, the agent uses their profiles to propose complementary layers, e.g., that include demographic or infrastructure information. 

To integrate these datasets, the \compOperationCopilot leverages the metadata to classify each variable's measurement type (i.e., extensive or intensive) and recommends valid operators, such as areal-weighted interpolation or sum aggregation, before any computation is performed. Finally, 
%at the synthesis stage, \name 
\revision{the LLM \compSynthesisAgent}
renders harmonized layers into a unified view, inferring directional polarity relative to the stated objective to produce an objective-specific priority map that visualizes areas of need and opportunity.
% JF: there was no mention of provenance, and in the contributions, the provenance appeared out of context
% \jf{I have added this sentence -- as provenance is mentioned in the contributions without having been explained}
\revision{To support transparency and user steering, the system captures detailed provenance of the process.}

 % \scq{Commenting out this paragraph since VIS papers usually list evaluation as a contribution, and it’s already included in the list of contributions. }
 
% \paragraph{Validation and Scope.} We evaluate the framework through a multi-tiered strategy using a custom benchmark of 28 urban research scenarios, constructed over a curated NYC OpenData lake containing 112 geospatial datasets, and 100 spatial mapping tasks. 
% \jf{using -- give brief description of the benchmark used} \scq{done}
% First, quantitative ablations confirm the importance of metadata profiling: \compAtlas-generated descriptions outperform baseline LLMs in dataset discovery across 28 scenarios, and \compOperationCopilot achieves 100\% semantic and 87\% geometric validity across 100 mapping tasks, preventing silent aggregation errors. Second, we demonstrate the effectiveness of \name through two real-world case studies, after-school prioritization and housing opportunity analysis. Finally, interviews with four domain experts (\textbf{E1--E4}) show that \name transforms the LLM into a collaborator, re-framing sensitivity to spatial aggregation and mapping decisions as an exploratory visual capability rather than a methodological burden.

 In summary, our main contributions can be summarized as follows:
\begin{itemize}
    \item We present \name, a novel visual analytics system that transforms \hide{opaque,} manual GIS data-wrangling into a transparent 
    %, node-based 
    workflow, enabling analysts to \hide{explicitly} navigate data discovery, spatial boundaries, and harmonization within a unified exploratory environment.
    \item We design a pipeline of \emph{context-aware AI agents} 
    %grounded in real-world data distributions, 
    including: an offline \compAtlas for metadata extraction, a \compDiscoveryAgent that retrieves datasets based on high-level analytical goals, and a \compOperationCopilot \revision{that enforces measurement semantics to prevent silent aggregation errors.}
    \item \revision{We design an \emph{integrated visual interface} where an \textit{Integration Provenance Graph}, \textit{Multivariate Priority Map}, and \textit{Spatial Delta Map} work in tandem to support structural transparency and counterfactual exploration across alternative spatial configurations.}
    %, enabling analysts to interactively track how \hide{high-level} conclusions shift across alternative boundary and semantic configurations.}
    % \item \revision{We design an \emph{integrated visual interface} where specialized components work in tandem to support structural transparency and counterfactual exploration of spatial data. This cohesive workspace coordinates an \textit{Integration Provenance Graph} for auditing the AI engine's underlying operator choices with tightly coupled \textit{Multivariate Priority} and \textit{Spatial Delta Maps}, enabling analysts to interactively track how high-level conclusions shift across alternative boundary and semantic configurations.}
    % \item We introduce \emph{interactive visual representations} to make spatial harmonization explicit and explorable, featuring \revision{an \textit{Integration Provenance Graph} that visualizes the exact operators applied by the AI engine, } a \textit{Multivariate Priority Map} that synthesizes layers to visualize areas of interest, \revision{and} a \textit{Spatial Delta Map} for exploring how conclusions shift across different configurations.
    %a \textit{Multivariate Priority Map} that synthesizes layers to visualize areas of interest, a \textit{Spatial Delta Map} for exploring how conclusions shift across different configurations, and an \textit{Integration Provenance Graph} that visualizes the exact operators applied by the AI engine.
    \item We conduct a \emph{comprehensive multi-tiered evaluation} of the framework, including quantitative ablation studies that establish the necessity of explicit statistical profiling for LLM spatial reasoning, two real-world urban planning case studies, and domain expert interviews demonstrating the system's ability to turn sensitivity to spatial aggregation and mapping operators from a methodological burden into an exploratory visual asset.
\end{itemize}

\section{Related Work}
\label{sec:related}

While urban open data initiatives provide broad access to raw spatial information~\cite{nycopendata, furmancenter2024, capitalplanning}, the \emph{last mile} of integration, reconciling geometries and harmonizing statistical semantics, remains a manual, error-prone endeavor. Existing GIS platforms and spatial libraries (e.g., \emph{ArcGIS}~\cite{arcgis}, \emph{Apache Sedona}~\cite{sedona_geospark}, \emph{GeoPandas}~\cite{geopandas}, \emph{PySAL}~\cite{Rey2010PySAL}) offer robust procedural foundations for spatial ETL. However, these workflows lack semantic awareness, placing the analytical burden entirely on users to manually identify relevant datasets and determine statistically valid operators (e.g., distinguishing \revision{\emph{intensive}} from \revision{\emph{extensive}} variables). While these tools can execute any specified integration, they support neither intent-driven discovery nor automated operator recommendation, capabilities crucial for urban decision-making. Overcoming these limitations requires solving interconnected challenges in profiling, search, spatial reasoning, and visual analytics\revision{, expanding upon emerging paradigms in LLM-supported urban planning~\cite{Zheng2025}}. In what follows, we review prior work that addressed these challenges.
%domains, albeit in a largely fragmented way.

\paragraph{Uncovering Data Semantics.}
A prerequisite for urban data integration is recovering the meaning of raw attributes before they can be searched, aligned, or transformed. \hide{In data management, t}\revision{T}his is commonly \revision{addressed} through column type annotation (CTA), which assigns semantic labels to table columns %and supports downstream tasks such as schema matching, search, and integration.
\revision{to support schema matching, search, and integration.}
%
% CTA has been studied extensively as a general table-understanding task. Representative systems include context-aware 
\revision{CTA approaches have been proposed that use different methods, including} learning-based  such as \emph{Sato}~\cite{zhang2020sato}, table representation models such as \emph{TURL}~\cite{deng2022turl}, and multi-task frameworks such as \emph{Doduo}~\cite{suhara2022doduo},
%that jointly model column semantics and inter-column relations. More recent work explored the use of LLMs for CTA~\cite{feuer2023archetype}. 
\revision{and more recently, LLM-based approaches}~\cite{feuer2023archetype}.
%\emph{ArcheType}~\cite{feuer2023archetype}. 
While these general approaches leverage table context and relational cues, work that focuses specifically on geospatial type detection remains limited and is often embedded within broader GIS or semantic-enrichment pipelines. Instead, existing work has focused on recovering spatial semantics from other raw forms, for example, by detecting geospatial location descriptions in natural language~\cite{stock2022detecting} or by converting raw GPS trajectories into semantic place- and route-based representations to support semantic querying~\cite{aldohuki2017semantictraj}. \name builds on the general CTA literature but targets a narrower and more operational goal: identifying, at ingestion time, the specific attribute types that can anchor a dataset in space, such as coordinates, administrative identifiers, address-like fields, and geometry encodings. \compAtlas specializes in semantic typing for geospatial attributes, rather than open-ended table understanding.

\paragraph{Mixed-Initiative Data Discovery.}
Spatial data discovery has traditionally been supported through geoportals and repository search interfaces, which help users locate datasets via metadata records, keyword search, and service catalogs. 
%While essential for publishing spatial resources, 
\revision{These} systems typically assume users already know what they are looking for and provide limited support for semantic intent matching. Recent studies continually highlight issues such as weak search support, limited guidance, and poor metadata quality in practical geodata discovery workflows~\cite{ankama2026geoportals,ferrari2024ogcsearch}, as well as the persistent difficulty of assembling heterogeneous spatial sources in practice
\cite{lemmens2015geospatialdiscovery,he2024oge}.
To move beyond raw filenames and manually created metadata, profiling has emerged in data management as a key mechanism to make large repositories searchable. Systems like %\emph{Data Civilizer}, \emph{Aurum}~\cite{deng2017thedatacivilizer,fernandez2018aurum}, and 
\emph{Auctus}~\cite{castelo2021auctus} \revision{have shown} how precomputed profiles,  sketches, and explicit spatial/temporal column detection can support dataset exploration and spatio-temporal search. \name builds on this profiling paradigm\revision{,} and 
%outperforms traditional metadata-dependent 
extends discovery workflows with geospatial column-type annotations and validation signals tailored to urban data.

\revision{Recent} research leverages LLMs to enhance data semantics and discovery capabilities. Frameworks like \emph{ArcheType}~\cite{feuer2023archetype} support retrieval-augmented CTA, while systems like \emph{AutoDDG}~\cite{zhang2025autoddg} leverages dataset contents to generate textual summaries and improve findability. Concurrently, GIS research has explored LLM-centered geospatial agents for \revision{executing natural-language tools and retrieving data}~\cite {zhang2024geogpt, ning2024autonomousgisagent}. While these efforts highlight the promise of intent-aware interaction, they typically assume the presence of external toolchains or high-quality metadata. \name overcomes these limitations by grounding discovery locally via ingestion-time profiling that produces compact, reliable spatial signatures. By using the \compAtlas to extract structured profiles as inputs for automated description generation, \name combines these structural signatures with generated summaries, enabling LLM agents to discover datasets that match a user's analytical intent.
%both structural spatial realities and the user's high-level analytical intent.

\paragraph{Spatial Operator Recommendation.}
Selecting appropriate mapping and aggregation operators is a fundamental challenge in multivariate spatial integration, as improper aggregation across spatial supports can introduce statistical bias and artifacts associated with \emph{MAUP}~\cite{goodchild1980areal,flowerdew1991areal,langford2006areal}. While classical measurement theory distinguishes between extensive (e.g., counts) and intensive (e.g., densities) variables to determine valid aggregation strategies~\cite{stevens1946measurement, goodchild2010geographic}, modern GIS platforms typically force analysts to manually specify these transformations. Attempts to automate this via ontology-based GIS rely heavily on manually curated schemas, limiting their scalability in practical workflows~\cite{Fonseca2002ontology, Top2022ontology, Tian2008dataqueryintegration}. While recent LLM-powered frameworks successfully translate natural-language queries into executable GIS pipelines\revision{~\cite{zhang2024geogpt,akinboyewa2025giscopilot,luo2025geojsonagents,zhang2025geoanalystbench}}, they largely treat spatial operators as predefined tools rather than reasoning about their statistical suitability. \name targets this gap by introducing an AI Copilot that recommends valid integration strategies using dataset metadata, geometric properties, and semantic reasoning. The reasoning is grounded by the semantic information extracted by \compAtlas, 
%which infers spatial column semantics directly from raw data, collectively 
which enables the automated, mathematically valid harmonization of heterogeneous spatial datasets\revision{, aligning with human-centric principles in GeoAI-enabled urban computing~\cite{lu2025geoai}.}
%\revision{, aligning with core objectives in human-centric GeoAI.}
%\revision{. This hybrid design addresses core challenges in human-centric GeoAI by anchoring automated reasoning to empirical data profiles~\cite{lu2025geoai}.}

\paragraph{Visual Interfaces for Data Integration.}
%Visual interfaces keep analysts in the loop during data integration, with tabular systems exposing 
%\revision{Visual interfaces play a vital role in data integration, translating complex, heterogeneous urban data for decision-making~\cite{deng2023survey}. While tabular systems expose} 
%candidate correspondences for human refinement~\cite{koutras2021valentine,wu2026bdiviz} and visual programming environments (e.g., \emph{VisTrails}, \emph{Orange}, \emph{KNIME}, \emph{GeoVISTA Studio}) using explicit graph structures to represent pipelines~\cite{bavoil2005vistrails,demsar2013orange,dietz2020knime,takatsuka2001geovistastudio}. 
\revision{To render complex, heterogeneous urban data actionable~\cite{deng2023survey}, visual interfaces keep analysts in the loop during data integration. Tabular systems expose candidate correspondences for human refinement~\cite{koutras2021valentine,wu2026bdiviz}, while visual programming environments (e.g., \emph{VisTrails}, \emph{Orange}, \emph{KNIME}, \emph{GeoVISTA Studio}) use explicit graph structures to represent pipelines~\cite{bavoil2005vistrails,demsar2013orange,dietz2020knime,takatsuka2001geovistastudio}.}
For spatial analytics, tools like \emph{VAUD} enable cross-domain urban querying over heterogeneous city data~\cite{chen2018vaud}, while systems like \emph{GeoViz Toolkit}, \emph{GeoDa Web}, and \emph{GOdIVA} provide coordinated views for web-based analytics and interactive semantic enrichment~\cite{hardisty2011geoviz,li2015geodaweb,ding2020godiva}. More recently,
%Recent spatial-analysis systems further extend this work to 
approaches have been proposed to support report authoring via contextualized explanations (\emph{GeoExplainer}~\cite{lei2024geoexplainer}) and multi-granularity map comparison (\emph{S-VIDIA}~\cite{bastos2023svidia}). 
Emerging LLM-based copilots like \emph{ChatGeoAI} and \emph{GIS Copilot} have begun integrating natural-language-driven analysis directly into GIS platforms~\cite{mansourian2024chatgeoai,akinboyewa2025giscopilot}. \name combines graph-structured workflows and copilot assistance but targets a distinct integration challenge: constructing pipelines to integrate datasets with disparate geometries and statistical semantics. Rather than simply inspecting spatial data or executing predefined queries, our interface is designed to help analysts discover relevant datasets, iteratively compare alternative harmonization strategies, and maintain transparent analytical provenance throughout the entire process.

\section{\name}
% \section{\textit{UrbanTrace}}
\label{sec:system_architecture}

\name is a visual analytics system designed to help analysts discover, integrate, and explore heterogeneous spatial datasets to solve complex spatial problems. By maintaining continuous dialogue between the human analyst and an ensemble of LLM-driven agents, \name streamlines spatial data discovery and harmonization. 
%\jf{streamlines (since the user is in the loop, the process is not automated)} the tedious and 
%\jf{why risky? challenging} mathematically risky steps of data harmonization. 

%\jf{this seems to be out of place here}
%To ensure transparency and trust, the system explicitly surfaces analytical provenance, mitigating the "black-box" issues often associated with AI-driven analysis, and provides synthesized multivariate maps to dynamically explore the integrated results.

\subsection{Domain Requirements}
\label{sec:domain_requirements}

Integrating heterogeneous spatial datasets introduces significant analytical and cognitive hurdles. \revision{Following design-study methodology~\cite{sedlmair2012}, we derived the following desiderata for \name from the core challenges (Sec.~1) and our year-long engagement in the OSCUR project~\cite{oscur2024}:}%, leaving the expert sessions (Sec.\ref{sec:expert_feedback}) as a strictly summative validation:}

%\jf{maybe combine r2 and r3 into semantic understanding; or in the response to reviewer 2, we can acknowledge that both requirements are related to understanding, and that we separate them for clarity; also R2 is also connected to C1 -- profiling generates informaiton that helps with discovery....}
%of spatial data science and MAUP-related artifacts, we distill the design of \name into five core desiderata (R1--R5):

%\jf{we need to distinguish the requirement from the solution: Data Understanding can be addressed by Semantic Profiling }
%\jf{These requirements must be properly motivated; this can be done using an example in the introduction, or if we want the introduction to be shorter, we could have a subsection 3.1 that describes the problem in detail and provides a segue into the requirements}
\begin{enumerate}[start=1,label={[\bfseries R\arabic*]}]
    \item \label{req:datadiscovery} \cdatadiscovery~\emph{Context-Aware Data Discovery}: Analysts often struggle to find relevant datasets for their analyses.
    %across disconnected data portals. 
    The system must dynamically recommend complementary spatial data based on the user task and the datasets currently in use.
        \item \label{req:profiler} \cdatadiscovery\cmeasurement~\emph{Metadata Generation}: \revision{Open datasets are often published with terse descriptions and insufficient metadata. To support data discovery and integration,      
        the system must automatically derive the necessary metadata in a data-driven fashion,  by inferring attributes' semantics as well as geometric and statistical properties directly from source files.}
    % \item \label{req:profiler} \cmeasurement~\emph{Automated Semantic Profiling}: Raw spatial data lacks the metadata required for \emph{correct} integration. The system must infer geometric constraints and semantics for the data (e.g., intensive vs. extensive variables) to prevent \revision{downstream aggregation errors.}
   \item \label{req:integration} \cspatialunit\cmeasurement~\emph{Semantic-Aware Spatial Harmonization}: Manually specifying spatial joins across unaligned geographic boundaries is error-prone. Rather than requiring users to be GIS experts, the system must guide analysts through the harmonization process by recommending valid aggregation pathways, ensuring, for example, that extensive counts are summed and intensive rates are area-weighted.
    % \item \label{req:integration} Mathematically Sound  \jf{why mathematically sound?} Harmonization: \jf{Should we say something more like "Guiding Users through Harmonization Process"} Manually specifying spatial joins and aggregation operators across unaligned geometries is highly error-prone. The system should automatically infer and recommend statistically \jf{we say mathematically, statistically, without explaining what this means} valid transformation pathways. \jf{we need to explain this in more detail. What is valid?} \joao{Do we need a "mathematical formalization" of data harmonization here? Are we using it to prove or demonstrate anything?}

    \item \label{req:explanableprovenance} \cdatadiscovery\cspatialunit\cmeasurement~\emph{Explainable Visual Provenance}: AI-assisted workflows risk becoming opaque "black boxes." The system must expose the complete analytical evidence chain, allowing users to interpret, audit, and manually override automated decisions.

    %\joao{why synthesis?}\scq{added a sentence explaining what synthesis is in our workflow}
    \item \label{req:multivariatesynthesis} \cspatialunit~\emph{Multivariate Synthesis 
    and Spatial Comparison}: Answering complex urban questions requires normalizing and combining multiple variables into a %single 
    \revision{unified} view (synthesis). Because these results are highly sensitive to zoning boundaries, the system must enable explicit visual comparison of how synthesized results shift when underlying spatial parameters or boundaries are modified. 
    %The system must support rapid exploration of these synthesized results and enable explicit spatial comparison (counterfactuals \joao{What are counterfactuals here? Is it the same as recourse?}) when analytical parameters are modified.
\vspace{-.15cm}
\end{enumerate}

%\subsection{System Architecture}
\subsection{System Overview}
\label{sec:pipeline}

\begin{figure}[t]
  \centering
  \includegraphics[width=1\linewidth]{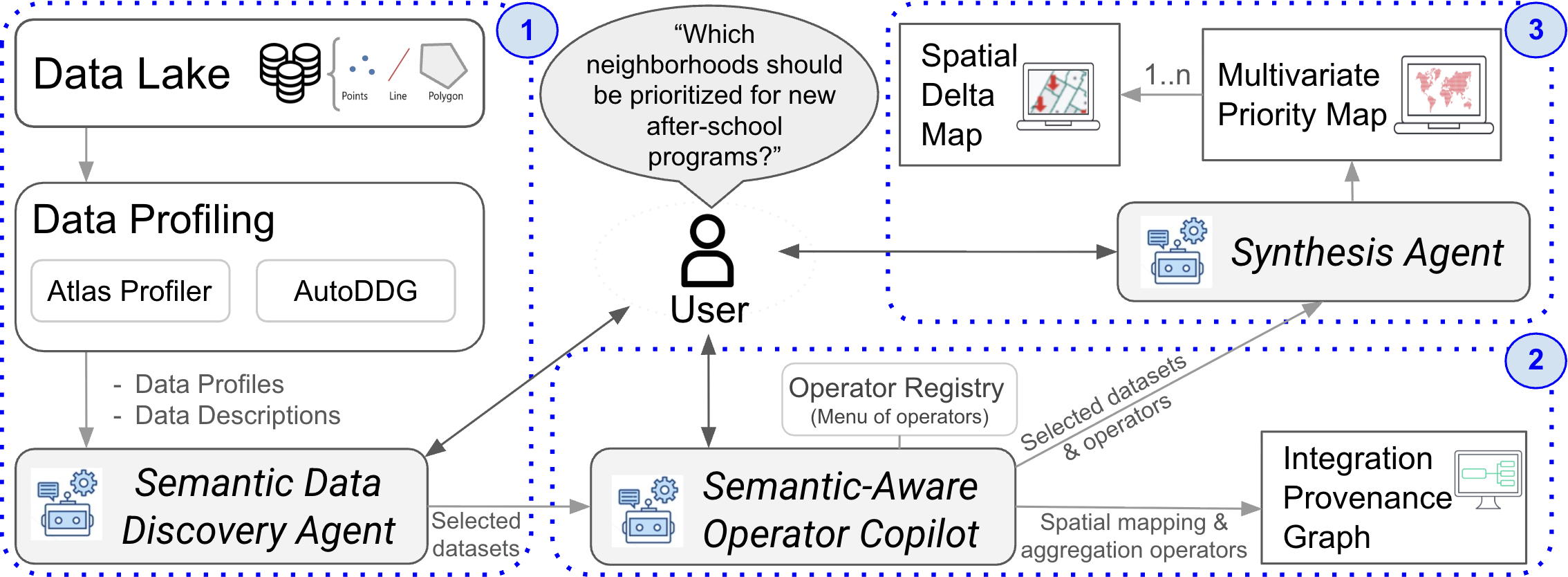}
  \vspace{-0.4cm}
  \caption{\revision{\name architecture.}}
  \label{fig:urbanTrace_architecture}
  \vspace{-0.6cm}
\end{figure}

To fulfill these requirements, \name employs a three-stage pipeline (Figure~\ref{fig:urbanTrace_architecture}) that users interactively refine using the \compCanvas (\autoref{sec:canvas_environment}). Offline, the \compAtlas (\autoref{sec:atlas_profiler}) extracts semantic and geometric metadata from datasets in a data lake (\ref{req:profiler}) which enables both data discovery and 
spatial reasoning.
% operation selection \jf{need to fix}. 
The workflow begins with intent-driven discovery (\ref{req:datadiscovery}): given a user-defined question or analysis description, the \revision{\compDiscoveryAgent} retrieves candidate datasets. Next, the \compOperationCopilot uses the extracted metadata to recommend 
% \jf{what is the definition of mathematically valid mapping? should we say just "valid"} 
valid spatial mapping and aggregation operators (\ref{req:integration})\revision{, while recording all operations in a transparent \emph{Integration Provenance Graph} (\ref{req:explanableprovenance}).}
% ensuring, for instance, that appropriate operators are chosen for intensive versus extensive variables (\ref{req:integration}).
Finally, an LLM \compSynthesisAgent compiles the harmonized data into an interpretable \emph{Multivariate Priority Map} (\ref{req:multivariatesynthesis})\revision{.}

% \begin{figure}[t]
%   \centering
%   \includegraphics[width=0.75\linewidth]{figs/urbanTrace_architecture_v3.png}
%   \caption{Overview of \name’s architecture.}
%   \label{fig:urbanTrace_architecture}
%   \vspace{-0.4cm}
% \end{figure}

\subsection{\revision{Analytical Foundations}}
\subsubsection{\revision{\compAtlas: Spatial Column Type Annotation}}
\label{sec:atlas_profiler}

A fundamental constraint on integrating LLMs into geospatial pipelines is that they cannot directly ingest raw spatial datasets due to strict context-window limits and the geometric complexity of vector data. To bridge this gap, \name introduces the \compAtlas, an offline hybrid learning- and rule-based pipeline that converts raw tabular data into compact spatial profiles at ingestion time \ref{req:profiler}.

\paragraph{Profile Construction and Corpus Generation.}
The profiler maps columns to a bounded vocabulary of operational geospatial types $\mathcal{T}$ (Table~\ref{tab:atlas_classes}). To train the classifier, we construct a corpus of human-annotated examples augmented by LLM-generated synthetic data. This augmentation helps distinguish surface regularities (e.g., \texttt{latitude}, \texttt{longitude}, \texttt{polygon}, \texttt{point}) from \emph{hard negatives}---columns with deceptively similar lexical forms (e.g., \texttt{borough\_code} vs. generic numeric identifiers, or \texttt{address} vs. free-form text).
\vspace{-0.7em}
\begin{table}[htbp]
\centering
\caption{Supported classes in the \compAtlas geospatial classifier.}
\small
% \begin{tabular}{lp{4.8cm}}
\begin{tabular}{lp{5.5cm}}
\toprule
\textbf{Category} & \textbf{Recognized classes} \\
\midrule
Geographic coordinates & \texttt{latitude}, \texttt{longitude} \\
Projected coordinates & \texttt{x\_coord}, \texttt{y\_coord} \\
Admin identifiers & \texttt{bbl}, \texttt{bin}, \texttt{zip\_code}, \texttt{borough\_code} \\
Location strings & \texttt{city}, \texttt{state}, \texttt{address} \\
WKT geometries & \texttt{point}, \texttt{line}, \texttt{polygon}, \texttt{multi\_polygon}, \texttt{multi\_line} \\
Other & \texttt{non\_spatial} \\
\bottomrule
\end{tabular}
\label{tab:atlas_classes}
\vspace{-.3cm}
\end{table}

% \jf{this is describing how we serialize columns; like magneto, it repeats the column names, but we are not explaining why...}
% \jf{did we try different serialization strategies? If not, maybe we can just say that we use the same strategy as magneto} 
\paragraph{Lexical and Semantic Embedding.}
For each column $c$, the profiler constructs a token sequence $S_c$ from its header and sampled values. Following the serialization strategy used in \emph{Magneto}~\cite{liu2024magneto}, we repeat the column name to increase its salience relative to the sampled values before encoding. Concretely, if $n_c$ is the column name and $V_c = \{v_1, \dots, v_k\}$ is a random sample of observed values, we serialize the column as
\vspace{-0.68em}
\begin{equation}
  S_c \;=\;
  \Bigl(\bigoplus_{j=1}^{\gamma} [\mathrm{COL}] \oplus n_c \Bigr)
  \oplus
  \Bigl(\bigoplus_{i=1}^{k} [\mathrm{VAL}] \oplus v_i \Bigr),
  \label{eq:token_seq}
\end{equation}
%\vspace{-0.1cm}
where $\oplus$ denotes concatenation and $[\mathrm{COL}]$, $[\mathrm{VAL}]$ are delimiter tokens. We encode $S_c$ using a \texttt{BAAI/bge-base-en-v1.5} model, which we fine-tune for this task with contrastive learning. The encoder outputs token-level hidden states, and we apply mean pooling to obtain a single 768-dimensional representation for the entire column.

% \jf{does atlas profile use fine-tuning of existing models? if so, which models are used? and how do you construct the training data?  what is the pre-defined set of target classes that the system supports?}
\paragraph{Contrastive Separation and Hybrid Classification.}
Since the target label space is predefined and operational, the main challenge is not open-ended semantic interpretation, but separating \emph{hard negatives}. Let $s(c,c')$ denote cosine similarity between normalized embeddings. For each column $c$, we mine hard negatives from its top-$k$ nearest neighbors $\mathcal{N}_k(c)$, selecting the $h$ most similar columns with different labels: $\mathcal{H}(c) = \{c' \in \mathcal{N}_k(c) \mid y_{c'} \neq y_c\}_{1:h}$.
These hard negatives guide a joint objective combining cross-entropy and supervised contrastive loss: $\mathcal{L}_{\mathrm{total}} = \alpha\,\mathcal{L}_{\mathrm{SupCon}} + (1 - \alpha)\,\mathcal{L}_{\mathrm{CE}}$, where $\mathcal{L}_{\mathrm{CE}}$ is cross-entropy, $\mathcal{L}_{\mathrm{SupCon}}$ is supervised contrastive loss \cite{khosla2020supcon}, and $\alpha$ balances the two. This objective pulls same-type columns together while separating hard negatives, improving classification over the target types $\mathcal{T}$.

\paragraph{Rule-Based Validation and Geometric Profiling.}
Neural predictions are verified by a rule-based stage enforcing physical constraints (e.g., latitude $\in [-90, 90]$). Beyond semantic typing, the profiler also computes spatial coverage (e.g., bounding boxes, temporal extents) and summary statistics (mean, variance, distinct counts). 
% \jf{what does this json look like? does it also include a description? Do you use AutoDDG?}
% The resulting compact JSON profile is stored for online retrieval, grounding the \jf{should be consistent with the operator name, maybe create a macro for it} Operator Copilot's subsequent reasoning in observed statistical realities and reducing the risk of analytical hallucination.
The result is stored as a compact JSON profile for later retrieval. This JSON contains structured metadata only;
\revision{without}
%it does not include 
a generated natural-language description. Instead, \compAtlas's output is passed downstream to the \autoDdg~\cite{zhang2025autoddg} description-generation step, which uses the structural profile to synthesize a dataset description. These stored profiles then ground the \compOperationCopilot's reasoning in observed dataset characteristics thus reducing the risk of analytical hallucination.

%% ===
%%  Formal Model of Strategy-Based Spatial Integration
%% ===

\subsubsection{\revision{Formal Model of Spatial Integration}}
\label{sec:formal_model}

To integrate heterogeneous datasets \revision{while reducing spatial aggregation errors}~\cite{Gotway2002combinespatialdata, WONG2009MAUP}, \name uses a deterministic geometric engine. Datasets ($D^{(k)}$) differ in geometry type (points, lines, polygons), resolution, and semantics, 
%\jf{I don't understand what you mean by "fixed spatial support"}
%\scq{by fixed spatial support, I mean, a common uniform grid to map all geometry types onto before performing any spatial operation. A better term would be "standardized reference grid". I will update this. OK}
so integration is defined over a standardized reference grid. Rather than directly joining unaligned datasets, which compounds spatial misalignment errors, 
% \jf{are these 'topological' errors?  or geometric mismatch errors?} \scq{what I actually mean are spatial misalignment issues, for example, assigning points to the wrong polygon because the boundaries don't perfectly line up when you try to directly intersect unaligned layers. I will update the text to "spatial misalignment errors" to be more precise.}
we define a common regular grid partition $S = \{c_1, \dots, c_M\}$ over the study region $\Omega \subset \mathbb{R}^2$, where each $c_j$ is a grid cell and $M$ \revision{is determined by the selected grid resolution (default = 8).} This geometry-agnostic intermediate layer isolates initial spatial mapping from downstream reporting units\revision{, allowing analysts to
examine sensitivity to MAUP across different spatial granularities.}

\paragraph{Dimensions of an Integration Strategy.}
We define a 
%complete 
spatial integration strategy for a dataset $D^{(k)}$ as a sequence of three operator classes: $\mathcal{I}^{(k)} = Z \circ A_1 \circ R$. First, a \textbf{Spatial Mapping Operator} ($R$) maps raw input geometries to grid cells by assigning geometric weights $w_{ij}$ (e.g., via binary containment, nearest assignment, or proportional intersection). Second, a \textbf{Cell-Level Aggregation Operator} ($A_1$) summarizes the mapped values within each cell $c_j$ (e.g., via additive totals or central tendencies). Finally, a \textbf{Zoning Operator} ($Z$) aggregates the intermediate grid cells into the user's final reporting units (e.g., Council Districts) using area-weighted overlap or centroid containment.

\paragraph{Constraining the Operator Space.}
Allowing an LLM to choose operators in an unconstrained fashion may lead to errors  (e.g., applying area-weighted mapping to dimensionless points). To ensure validity, we constrain the admissible families of operators based on the dataset's profile. \textbf{Geometric constraints} bound $R$: point geometries are restricted to proximity or containment families, while polygons are strictly routed through proportional area-weighted operators to ensure mass preservation\revision{~\cite{goodchild1980areal}}. \textbf{Semantic constraints} bound $A_1$ and $Z$ \cite{Chrisman01011998}: extensive variables (e.g., population counts) permit \emph{Sum} aggregations, intensive rates (e.g., poverty percentages) strictly require \emph{Weighted Means}, and ordinal classifications default to \emph{Majority} logic. Enforcing these rules reduces the combinatorial strategy space to a finite, deterministic set of sound pathways, serving as a guardrail for the LLM-driven Copilot.

\subsection{\revision{LLM Agents}}
%% 
%%  Semantic Data Discovery Agent
%% 

\subsubsection{\revision{\compDiscoveryAgent}}
\label{sec:discovery_agent}

Open urban data portals support data discovery through keyword search, matching query terms to dataset metadata. Unless users know the correct terms that match the terminology used in the metadata, they are unlikely to discover relevant datasets. Besides, metadata is often incomplete and sometimes inconsistent with the data~\cite{zhang2025autoddg}. 
\name addresses this problem through the \compDiscoveryAgent \ref{req:datadiscovery}.
%, an LLM-driven retrieval mechanism \Ref{req:datadiscovery}. 
When an analyst provides a natural-language objective (e.g., \emph{``Identify neighborhoods underserved by after-school technology programs''}) or drops an initial dataset onto the canvas, the agent translates this intent into a curated set of spatial data recommendations. Rather than querying raw tables, the agent reasons over the compact dataset descriptions and structured metadata generated offline by \compAtlas and \autoDdg. These include geometric type, attribute keywords, column summaries, geospatial classifier outputs, and dataset-level textual descriptions.

\paragraph{Context-Aware Prompting and Complementary Retrieval.}
To generate relevant recommendations, the agent evaluates a tripartite prompt containing: (i) task-level system instructions, (ii) the localized data lake catalog (enumerating dataset identifiers, column profiles, and semantic descriptions), and (iii) a live serialization of the current canvas state, including active nodes, spatial connections, and visual encodings. 

% \jf{do you have experiments comparing the recommendations using just 1 versus 1+2 versus 1+2+3?} \jf{if there are lots of datasets, won't this fill up the context window? And even if it doesn't, this can be very expensive. How big are these prompts?} \ew{The ablation study is in Section \ref{sec:discovery_ablation}. And you're right, the prompts now are relatively large around 50k tokens, do we need to mention that we would like to have a retrieving method while passing the catelogs?}
By conditioning retrieval on the active analytical topology rather than on the initial text prompt alone, the agent moves beyond simple lexical matching to propose \emph{complementary} data layers. For instance, starting with a \emph{School Locations} dataset, it may suggest demographic (poverty rates), service-demand (youth population), or infrastructure (broadband) layers, necessary to complete the spatial workflow. 

To maintain user agency, these recommendations are surfaced as suggestions, each accompanied by proposed visual encodings and a natural-language justification (Figure~\ref{fig:discovery_canvas}). The system only includes data 
upon explicit user acceptance, preserving a mixed-initiative workflow where the agent offers contextually useful suggestions while the analyst retains control over which layers are incorporated into the analysis.

\subsubsection{\revision{\compOperationCopilot}}
\label{sec:operator_recommendation}

A critical challenge in multivariate spatial integration is \revision{selecting appropriate} mapping and aggregation operators. While the formal model established in Section~\ref{sec:formal_model} defines a deterministic space of valid integration strategies, manually navigating this space remains a daunting task for analysts. Na\"{i}ve selections, such as directly summing intensive variables like poverty rates, lead to severe spatial aggregation errors and scale-dependent distortions in spatial analysis~\cite{WONG2009MAUP}. To bridge this gap, \name introduces the \compOperationCopilot, an LLM-driven engine that automatically navigates these guardrails to recommend suitable operators \ref{req:integration}.

\paragraph{Semantic Classification and Operator Logic.}
Each source variable $v$ is characterized by a spatial signature $M_v = \{ \eta, \mu, D, S, G_{src} \}$ derived from the dataset profiles generated by the \compAtlas, where $\eta$ is the column name, $\mu$ is the mean, $D$ is the distinct value count, $S$ is a data sample, and $G_{src}$ is the source geometry. The Copilot abstracts the grid engine by deriving a mapping function $f: (M_v, G_{tgt}) \rightarrow (\mathcal{M}, \mathcal{A})$, where $\mathcal{M}$ defines the spatial mapping ($R, Z$) and $\mathcal{A}$ defines semantic aggregation ($A_1$) into the target geometry $G_{tgt}$.

To constrain the LLM's recommendation space, the Copilot first classifies the variable's semantic type $C(v)$ based on structural types in $S$ and lexical cues in $\eta$ (e.g., ``density'' implying intensive quantities). We define three classes based on foundational cartographic and geocomputational principles \cite{Scheider2019extensiveintensivepropertie, goodchild1980areal, Chrisman01011998}: \textbf{Extensive} ($C_{\text{ext}}$) for spatially dependent magnitudes that scale proportionally with area (e.g., population); \textbf{Intensive} ($C_{\text{int}}$) for spatially independent ratios or densities (e.g., poverty rate) that do not scale additively with area; and \textbf{Ordinal} ($C_{\text{ord}}$) for qualitative categories (e.g., zoning). Formally, we define $C(v) = C_{\text{ord}}$ if $D \le 10 \land S \subset \mathbb{Z}$; $C_{\text{int}}$ if $S \subset \mathbb{R} \setminus \mathbb{Z} \lor \text{is\_rate}(\eta)$; and $C_{\text{ext}}$ otherwise. The primary recommendation engine leverages an LLM routed via an AI Gateway. Prompted with $M_v$, $G_{tgt}$, and a zero-shot classification schema, the LLM determines the statistically valid operators based on the inferred $C(v)$.

\paragraph{Integration Provenance.}
To ensure the Copilot's automated decisions remain transparent, the system externalizes the analytical provenance through an interactive \revision{\emph{Integration Provenance} Graph} \ref{req:explanableprovenance}. This panel uses a semantically typed node-link grammar to visualize operator composition: rectangular cards for data states, triangles for mapping ($\mathcal{M}$), and circles for aggregation ($\mathcal{A}$), as shown in Figure~\ref{fig:teaser}(D). When the Copilot assigns an operator, it populates this graph using progressive disclosure. Hovering reveals an \emph{Operator Tooltip} with the LLM's parsed, natural-language rationale (e.g., \emph{``Poverty rate is classified as an intensive variable that does not scale with area; applying areal-weighted interpolation...''}). While exposing the complete step-by-step pipeline introduces a slight learning curve compared to a minimal ``black-box'' UI, this explicit provenance is critical. It enables analysts to trace data flows, verify AI-driven physical changes via coordinated map highlighting, audit mathematical guardrails, and manually override recommendations, ensuring that final outputs are trustworthy for decision-critical urban governance.

%% 
%%  Explainable Priority Synthesis
%% 

\subsubsection{\revision{Multivariate Synthesis and Counterfactuals}}
\label{sec:priority_synthesis}

With the diverse spatial datasets harmonized into a common zoning framework (Sections~\ref{sec:formal_model} and \ref{sec:operator_recommendation}), the final analytical challenge is translating these disparate indicators into a cohesive view. \name accomplishes this through the \textit{LLM Synthesis Agent}, which evaluates the harmonized variables against the user's original natural-language objective to infer semantic relationships and compute a final multivariate priority score (Req.~\ref{req:multivariatesynthesis}).

\paragraph{Directional Polarity and Weight Inference.}
In multivariate urban analysis, variables lack inherent positive or negative utility; their impact is strictly contextual. Let $\mathcal{O}$ be the user's stated analytical objective (e.g., \emph{``identify neighborhoods underserved by after-school technology programs''}). For each harmonized target variable $k$, the \revision{\compSynthesisAgent} evaluates its semantic relationship to $\mathcal{O}$ to infer two critical parameters. First, for \textit{Directional Polarity} ($d_k \in \{1, -1\}$), the LLM dynamically infers this orientation based on the specific goal. For example, if $\mathcal{O}$ targets areas with \emph{low} poverty, the poverty rate is assigned $d_k = -1$ (Inverted), meaning lower raw values yield a higher priority score. If no explicit objective is defined, the system defaults to a vulnerability heuristic: variables that correlate with a deficit or risk (e.g., poverty) are assigned $d_k = 1$ (Normal), whereas existing assets (e.g., broadband) are assigned $d_k = -1$. Second, \textit{Relative Importance} ($w_k$): the LLM assesses how directly tied the variable is to $\mathcal{O}$, assigning weights that the system normalizes to ensure $\sum_{k=1}^{K} w_k = 1$.

\paragraph{Multivariate Priority Scoring.}
\revision{Grounding this workflow in data-driven Multi-Criterion Decision Making (MCDM)~\cite{Huchang2023mcdm}, the system computes a composite index $P_\ell$ for each administrative zone $z_\ell$ by combining the agent's weights and polarities: $P_\ell = \sum_{k=1}^{K} w_k \cdot \phi(\hat{y}_\ell^{(k)}, d_k)$, where $\hat{y}_\ell^{(k)} \in [0, 1]$ is the min-max normalized variable, and the inversion function $\phi(x, 1) = x$, $\phi(x, -1) = 1 - x$ mathematically flips values to align with the urban goal $\mathcal{O}$. }
%\jf{What do you mean by "LLM service degrades"? And how do you detect this?} %\scq{I mean, when we can not use the LLM, for example, because of timeout or connection issues. When we can not connect to the LLM, we get an error, so we trigger the heuristic fallback.}
% To handle LLM failures and timeouts, a heuristic fallback defaults to keyword-based polarity routing and equal weighting ($w_k = 1/K$). 
% \jf{what is keyword-based polarity routing?} \scq{it is a simple predefined dictionary to assign +1 (normal) or -1(inverted) polarity based on the column name, for example, inverting a variable that contains "income" or "capacity". Should we maybe remove this? This was built to ensure the system doesn't break in case we don't have access to the LLM for some issues, like connections.  }
% \jf{This is an implementation detail and since it is not critical, or at least we are not making a case that it is, maybe it would be better to remove it} \scq{I agree. THanks!}
%
Users can manually override these weights via interactive sliders within the \emph{Hotspot Priority} legend (see bottom-right in Figure~\ref{fig:hotspot_combined} (Left)), enabling domain experts to dynamically test alternative analytical scenarios \revision{via real-time recalculation.}
The analytical output of this synthesis is the \textit{Multivariate Priority Map} \revision{(see Figure~\ref{fig:teaser} (B3))}, a choropleth map that highlights intervention hotspots. %Within the system's node-link interface, this map is displayed in a \textsf{Result Map Node} (see Figure~\ref{fig:teaser} (B3)). 
% \jf{fig 6 has not top -- is this the correct figure?} \scq{actually, since we are referring to the node-link interface, it is better to cite the Result Map Node that we have in the teaser. Updated. thanks!}
%To avoid presenting this as a black-box oracle, 
The interface shows the assigned weights, polarities, and the LLM's natural-language rationale via interactive tooltips (Figure \ref{fig:spatial_delta} (left)), ensuring that the final output remains fully traceable. 
%to the initial objective.

\paragraph{Counterfactual Analysis via the Spatial Delta Map.}
To support \revision{counterfactual scenario exploration,} \name provides a \revision{\emph{Spatial Delta Map}}~\ref{req:multivariatesynthesis}. \revision{By accepting two \textit{Multivariate Priority Maps} as inputs} (e.g., a baseline model and a modified counterfactual model), as shown in Figure~\ref{fig:teaser} (B4), the system automatically aligns spatial geometries and computes the exact signed difference ($\Delta = P_{\text{model 2}} - P_{\text{model 1}}$) for each shared unit. \revision{To mitigate occlusion and visual clutter typical of multi-channel map overlays~\cite{McNabb2019GlyphPlacement}, the absolute shift magnitude maps to a high-contrast sequential background color scale,} while spatial variance is encoded using toggleable directional glyphs (arrows). \revision{Informed by cartographic placement and layout strategies to preserve underlying color legibility~\cite{McNabb2019GlyphPlacement}, these glyphs are scaled} proportionally across discrete semantic tiers (Minor, Moderate, Significant, Major) \revision{and colored divergently} (green for priority increases, red for decreases, gray for negligible variance), \revision{immediately drawing the analyst's attention to the largest localized fluctuations.}

%\paragraph{Counterfactual Analysis via the Spatial Delta Map.}
%To support the exploration of multiple scenarios testing, \name provides a \textsf{Spatial Delta Map}~\ref{req:multivariatesynthesis}. Analysts often need to understand how the introduction of a new variable (e.g., housing cost) or the manual adjustment of a weight alters the spatial priority distribution. Accepting two \textsf{Result Map Nodes} as inputs (e.g., a baseline model and a modified counterfactual model), as shown in Figure~\ref{fig:teaser} (B4), the system automatically aligns the spatial geometries and computes the exact signed difference ($\Delta = P_{\text{model 2}} - P_{\text{model 1}}$) for each shared unit. To avoid visual clutter, it employs a multi-channel encoding: absolute shift magnitude maps to a high-contrast sequential background color scale, while spatial variance is encoded using toggleable directional glyphs (arrows). By scaling glyph size proportionally across discrete semantic tiers (Minor, Moderate, Significant, Major) and pairing them with divergent coloring (green for priority increases, red for decreases, gray for negligible variance), the system immediately draws the analyst's attention to areas experiencing the largest fluctuations, allowing rapid assessment of localized spatial impacts.

\subsection{\revision{\compCanvas: A Unified Visual Workspace}}
\label{sec:canvas_environment}
\revision{The following describes how these components are unified within a single interactive workspace.}
To support the iterative nature of spatial exploration, the primary user interface is an infinite, node-based spatial canvas,  shown in Figure~\ref{fig:teaser}(B). This flexible visual programming paradigm serves as the foundational workspace for discovery, harmonization, and synthesis.

% \jf{Why do you call the panel "integration topology"? what about "integration provenance"?}
% \scq{"integration provenance" works better since it represents the operator's provenance for workflow}
\paragraph{Node-Based Visual Provenance.}
The analytical state is externalized via a directed graph of interactive components (Figure ~\ref{fig:teaser} B). Source datasets are loaded as \emph{Dataset Nodes} (B1) (which map viewports to raw geometries), and users link them to \emph{Integration Nodes} (B2) that encapsulate the transformation logic. The integrated data is then 
%fed into the \emph{Result Map Node} (B3), which 
\revision{visualized in} the synthesized \emph{Multivariate Priority Map} (B3). 
The \revision{\emph{Integration Provenance} Graph} (Fig.~\ref{fig:teaser} D) explicitly shows the provenance of the results: the flow of data from raw inputs through the specific operations applied by the AI-assisted geometric engine (\ref{req:explanableprovenance}). 

\paragraph{Coordinated Spatial Synchronization.}
To support comparative reasoning across heterogeneous geometries, all map-bearing nodes employ a Coordinated Multiple Views (CMV) architecture. Spatial interactions, such as panning or zooming on a hotspot, automatically synchronize across all upstream viewports, allowing users to pre-attentively verify local distributions without losing spatial context.

\paragraph{Comparative Exploration and the Delta Map.}
The infinite canvas supports branching analyses (\ref{req:multivariatesynthesis}). Analysts can duplicate a pipeline, adjust weights or spatial operators, and evaluate alternative models.
To compare the results from different approaches, \name includes a \emph{Spatial Delta Map} 
%node 
(Fig.~\ref{fig:teaser} B4). Given two results, it 
%automatically 
computes and visualizes the exact geographic shifts (increases or decreases in priority) between competing models, thereby drawing the analyst's attention to areas exhibiting higher variance under altered assumptions.

%% ============================================================
%%  Evaluation Section
%%  Contains: Case Study 1 (NYC Universal After-School Initiative)
%%            [Case Study 2 placeholder]
%% ============================================================

% \section{Evaluation}
% \label{sec:evaluation}

% We evaluate our AI-Assisted Spatial Analytics framework through two case studies
% drawn from real urban policy contexts. Each study exercises a different facet of
% the pipeline: automated data harmonisation, AI-guided operator recommendation,
% and the generation of explainable priority maps that can be validated against
% ground-truth outcomes.

\vspace{-.3cm}
\section{Evaluation}
\label{sec:evaluation}

Our evaluation uses a multi-pronged approach to answer the following questions: 1) Do AI agents make accurate recommendations? 2) Does the system support effective real-world analyses? 3) Do domain experts find it useful in practice?
%
%We answer these questions through a multi-pronged strategy. 
First, we quantitatively assess the accuracy of the AI agents in two ablation studies: one targeting dataset discovery (Sec.~\ref{sec:discovery_ablation}) and one targeting spatial operator recommendation (Sec.~\ref{sec:ablation}). Second, we demonstrate end-to-end scenario exploration and priority map synthesis through two real-world case studies (Sec.~\ref{sec:cs1}--\ref{sec:cs2}). Third, we validate the system's practical utility and workflow integration through structured interviews with four domain experts (Sec.~\ref{sec:expert_feedback}).

\hide{We carried out a multi-pronged evaluation of \name. We quantitatively assessed 
%to assess: (1) 
%two quantitative ablations assessing 
the AI agents' accuracy in dataset discovery (Sec.~\ref{sec:discovery_ablation}) and spatial operator recommendation (Sec.~\ref{sec:ablation}).
We (2) two real-world case studies demonstrating end-to-end scenario exploration and priority map synthesis (Sec.~\ref{sec:cs1}--\ref{sec:cs2}); and (3) domain expert interviews validating the system's practical utility and workflow integration (Sec.~\ref{sec:expert_feedback}).}

%%%%%%%%%%%%%%%%%%%%%%%%
%%%%%%%%
%%%%%%%%%%%%%%

\vspace{-.2cm}

\subsection{\cdatadiscovery~Quantitative Ablation: Dataset Discovery }
% Performance -- removed to avoid extra line
\label{sec:discovery_ablation}

% Identifying which datasets are relevant to a given analytical goal is the first and most open-ended challenge \name addresses. 
Identifying relevant datasets for a given analytical goal is the first and most open-ended challenge \name addresses.
We designed a controlled ablation study to assess how effectively the \compDiscoveryAgent meets this challenge, and specifically whether the structured metadata generated offline by \compAtlas and the dataset descriptions produced by \autoDdg provide meaningful retrieval gains beyond what an LLM can infer from dataset names alone.
% \jf{Why use just the names? Don't these datasets come with a description? If they do, reviewers may argue that the comparison is not fair.}

\hide{
We designed a controlled ablation study to assess the effectiveness of the \compDiscoveryAgent in retrieving 
%policy-relevant datasets 
relevant datasets from the \name data lake. The goal of this study is to evaluate whether the agent benefits from the structured metadata generated offline by \compAtlas and the dataset descriptions generated by \autoDdg, beyond what can be inferred from dataset names alone.}

\paragraph{Benchmark Tasks and Data Lake Scope.}
We created a benchmark of \textbf{28 real-world urban research scenarios} constructed using a curated NYC OpenData lake
containing \textbf{112 geospatial datasets}. For each dataset, we precompute structured profiles using \compAtlas and dataset-level descriptions using \autoDdg. The benchmark tasks span multiple policy themes, including housing, education, mobility, environmental justice, public services, and composite neighborhood prioritization. Each task consists of: (i) 
%a natural-language prompt of approximately 100 words 
\revision{a \(\sim\)100-word prompt} describing a realistic urban research or planning goal; (ii) a manually validated set of \textbf{3 to 5 relevant datasets}; and (iii) supporting references grounding the scenario in real research, report, or policy. 
Note that the prompts describe the analytical goal without explicitly naming the correct datasets, so the benchmark evaluates semantic discovery rather than keyword matching.
% \jf{it would be useful to give an example; also, can we make the benchmark available so that the reviewers can examine it?}
For example, mapping NYC districts where low student performance and high crime overlap~\cite {laurito2019schoolclimate} requires the agent to choose the aligned education and community-district crime-rate datasets.

% For example, one benchmark asks the agent to identify NYC community districts where weaker student performance overlaps with higher crime burden~\cite{laurito2019schoolclimate}. A strong answer should choose the aligned education and community-district crime-rate datasets and treat the result as a screening layer rather than a causal claim.

\paragraph{Task Construction and Annotation.}
We adopted an LLM-assisted, human-validated workflow to construct the benchmark. 
The LLM first identified candidate research scenarios from multiple trusted public sources \hide{(e.g., NYC government sites, State Comptroller, NYC 311)} and proposed task formulations, along with plausible, relevant datasets from our catalog.
% the LLM first identified candidate research scenarios from multiple trusted public sources, including NYC government websites, the New York State Office of the State Comptroller, and NYC 311. Based on these source materials and the local data catalog, it then proposed candidate task formulations and plausible relevant datasets.
We manually curated each task, rewrote prompts to control realism and ambiguity, and validated the expected datasets against the contents of the \name repository. This process ensured that the benchmark reflects realistic discovery conditions in which the agent must connect a high-level analysis objective to appropriate data layers, while avoiding trivial name-based retrieval.

\begin{figure}[t]
\vspace{-0.2cm}
  \centering
  \includegraphics[width=0.8\linewidth]{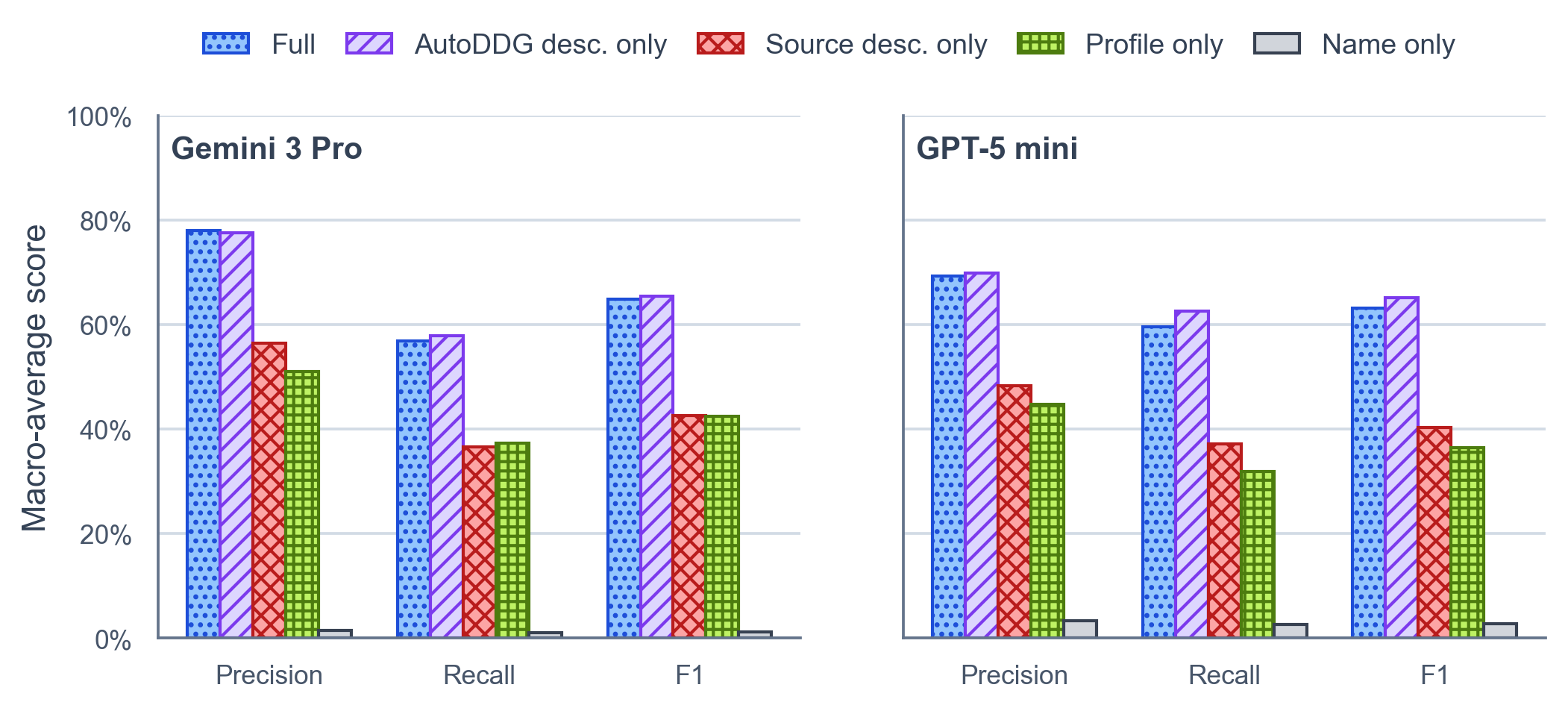}
  \caption{\revision{Macro-average precision, recall, and F1 for the dataset discovery ablation across 28 benchmark cases, using \texttt{Gemini 3 Pro} and \texttt{GPT-5 mini} as discovery backbones and averaging over five runs per case.}}
    \vspace{-0.5cm}
  \label{fig:discovery_ablation}
  %\vspace{-0.4cm}
\end{figure}

\paragraph{Baselines and Conditions.}
\hide{
For each benchmark task, the \compDiscoveryAgent receives the task prompt together
with dataset metadata from the local data lake, and returns a set of
recommended dataset identifiers. We evaluate four conditions as shown in Fig.~\ref{fig:discovery_ablation}: 
% \jf{do these datasets come with descriptions? If so, it would be interesting to include  "original description" in the ablation.} \ew{I collect the raw data from Nexus: Correlation Discovery over Collections of Spatio-Temporal Tabular Data, their dataset does not come with description, but the original NYC opendata entries does come with descriptions, the same goes to the datasets Sonia manually curated.}
(i)~\emph{Names Only}, in which the agent receives only dataset names;
(ii)~\emph{No profile} (Names + Descriptions), in which names are augmented with
\autoDdg-generated dataset descriptions;
(iii)~\emph{No description} (Names + Profiles), in which names are augmented with \compAtlas metadata; and
(iv)~\emph{Full}, which corresponds to the full
\name discovery configuration.
This comparison isolates the contribution of lexical metadata, structured profiles, and dataset-level semantic descriptions to urban dataset discovery.
}
For each benchmark task, the \compDiscoveryAgent receives the task prompt together with dataset metadata from the local data lake, and returns a set of recommended dataset identifiers. \revision{We evaluate five conditions as shown in Fig.~\ref{fig:discovery_ablation}: (i)~\emph{Name only}, in which the agent receives only dataset names; (ii)~\emph{Source description only}, in which names are augmented with original source descriptions from NYC OpenData; (iii)~\emph{Profile only}, in which names are augmented with explicit \compAtlas profile metadata; (iv)~\emph{\autoDdg description only}, in which names are augmented with \autoDdg-generated descriptions derived from \compAtlas profiles, but no explicit profile metadata is provided at inference time; and (v)~\emph{Full}, which combines \autoDdg-generated profile-derived descriptions with explicit \compAtlas profile metadata. The \emph{Full} condition does not include the original source descriptions. Dataset names are included in all conditions as they provide the reference handle needed for the agent to identify and return candidate datasets.}
This comparison isolates the contribution of lexical metadata, source-authored descriptions, structured profiles, and profile-derived semantic descriptions to urban dataset discovery.

\paragraph{Evaluation Protocol.}
To assess robustness across %different 
model families, we instantiate the \compDiscoveryAgent with two LLMs: \texttt{Gemini 3 Pro} and \texttt{GPT-5 mini}. For each combination of task, model, and ablation condition, we repeat the discovery process \textbf{5 times} to account for \revision{stochastic LLM variability. }%stochastic variance in LLM outputs. 
% 
%We evaluate retrieval quality 
\revision{Retrieval quality is evaluated} by comparing the set of datasets recommended by the agent against the manually validated ground-truth\hide{ set}, computing \emph{precision}, \emph{recall}, and \emph{F1} based on set overlap, and aggregating results across repeated runs and benchmark tasks. Note that raw accuracy over all datasets would be dominated by true negatives---most datasets in the lake are irrelevant to any given task---so these retrieval-oriented metrics more directly capture the agent's ability to surface what matters. Precision measures whether recommended datasets are relevant; recall measures whether relevant datasets are found; F1 balances the two. 
%
% This protocol evaluates whether richer metadata grounding improves the agent's ability to recover the datasets necessary to support a plausible downstream urban analysis, and whether that improvement is consistent across model families.
\revision{This protocol evaluates whether richer metadata grounding consistently improves dataset retrieval across model families, supporting downstream urban analysis.}

\hide{
We then compare the set of datasets recommended by the agent against the manually validated relevant set and compute \emph{precision}, \emph{recall}, and \emph{F1} over set overlap, aggregating the results across repeated runs and benchmark cases. Because each task contains only a small number of relevant datasets relative to the full data lake, these metrics more directly capture discovery quality than raw accuracy over all datasets, which would be dominated by true negatives. This protocol evaluates whether the agent can recover the datasets necessary to support a plausible downstream urban analysis.}

% \begin{figure}[htbp]

As shown in Fig.~\ref{fig:discovery_ablation}, \revision{both \emph{Full} and \emph{\autoDdg description only} achieve strong retrieval performance across models. The biggest gains come from profile-derived \autoDdg\ descriptions: for \texttt{Gemini~3~Pro}, F1 increases from 0.426 with source descriptions to 0.654 with \autoDdg descriptions; for \texttt{GPT-5~mini}, F1 increases from 0.403 to 0.651. This shows that \compAtlas\ contributes substantial value when its structured profiles are used by \autoDdg\ to generate dataset descriptions. Explicit profiles alone also improve substantially over \emph{Name only}, but they do not match the performance of profile-derived natural-language descriptions. This suggests that LLM-based discovery benefits most when structured profile metadata is translated into compact dataset-level semantic descriptions. Adding explicit profiles on top of \autoDdg descriptions does not consistently improve over \autoDdg descriptions alone, suggesting partial redundancy between the generated descriptions and the structured profile metadata.} At the other extreme, the \emph{Name only} condition performs near zero across all metrics, confirming that dataset names alone provide almost no useful signal for realistic urban data discovery.

\hide{The clearest pattern is the importance of the \autoDdg-generated description: removing descriptions causes a substantial degradation for both models, reducing F1 from \(0.648\) to \(0.424\) for \texttt{Gemini 3 Pro}, and from \(0.632\) to \(0.365\) for \texttt{GPT-5 mini}. In contrast, the \emph{Name Only} condition performs near zero in all metrics, confirming that dataset names alone provide almost no useful signal for realistic urban data discovery. }

%the profile information is  descriptions are themselves generated from \compAtlas-derived profiles; thus, the profiler's contribution remains present upstream, but is 
%
%mediated through the semantic description rather than through direct exposure of raw profile metadata to the LLM.

% ------------------------------------------------------------------
\vspace{-0.15cm}
\subsection{\cspatialunit\cmeasurement~Quantitative Ablation: Operator Selection }
% Accuracy % JF keep a single line
\label{sec:ablation}

We designed a second controlled experiment to assess the automated operator selection, benchmarking the Copilot across two dimensions corresponding to the framework's core logic: \emph{Geometric Validity} (the correct selection of the \textbf{spatial mapping operator} based on source and target geometries) and \emph{Semantic Validity} (the correct selection of the \textbf{aggregation operator} based on variable semantics, e.g., summing extensive absolute counts vs.\ averaging intensive rates). 

%46 unique column names were excluded, like IDs, coordinates.
\paragraph{Dataset Scope and Ground Truth.}
Our evaluation corpus comprises 25 real-world spatial datasets drawn from the 112-dataset collection described in Section~\ref{sec:discovery_ablation}: 17 MultiPolygon, 7 Point, and 1 MultiLineString, 
a distribution reflecting how municipalities aggregate sensitive data into polygons for geoprivacy~\cite{VanWey2005geoprivacy, furmancenter2024}.
% a distribution that reflects the polygon-dominated character of typical municipal open data ecosystems.  
% \jf{do you have statistics to back up this statement? that most datasets use polygons?} \scq{I do not have statistics. I wrote that because when collecting the datasets, almost all the datasets containing meaningful social variables (poverty, population, health) were bound to polygons. in practice, municipalities aggregate sensitive point-data (like health or poverty) into polygons for privacy reasons. should we just say then " a distribution reflecting how municipal open data routinely aggregates sensitive information into areal units (polygons) to preserve privacy." or should we just remove it? }
After excluding uninformative fields such as identifiers and coordinate columns, the corpus contains 215 semantically meaningful target columns.
Variable semantics are context-dependent (e.g., a column labeled \texttt{2023} can represent absolute counts or intensive rates); therefore, we annotated semantics at the dataset--column level. By decomposing the benchmark of 28 complex urban research scenarios into their pairwise dataset integrations, we extracted exactly \textbf{100 distinct spatial mapping scenarios}. Finally, to establish ground truth, a semi-automatic workflow (LLM preliminary classification followed by manual curation) classified the relevant variables within these 100 scenarios as extensive, intensive, or ordinal.

% A semi-automatic workflow (LLM preliminary classification followed by independent expert verification) classified each variable as extensive, intensive, or ordinal, systematically yielding \textbf{100 spatial mapping scenarios} for the study.

\paragraph{Baselines and Conditions.}
To test the necessity of statistical context, we evaluated five conditions: (i)~a \emph{Rule-Based} legacy GIS baseline; (ii--iv)~three \emph{LLMs} (GPT-4o-Mini, Claude~3.5~Sonnet, Claude~4.6~Opus); and (v)~our \compOperationCopilot. Following a strict protocol, conditions (i--iv) were provided with only the \textit{dataset names} (source and target) and \textit{column names}. They were explicitly denied access to data profiles (e.g., geometry types, data distributions) and did not perform our intermediate semantic classification step ($C(v)$), forcing them to infer both spatial and statistical properties from lexical cues alone.

% Notably, following a strict protocol, conditions (i--iv) were provided with only the \textit{dataset names} (source and target) and \textit{column names}. They were denied access to explicit geometry-type metadata or data distributions that we can get from a profiler, forcing them to infer both spatial and statistical properties from lexical cues alone.

\paragraph{Results and Trade-offs.}
Table~\ref{tab:results} summarizes the results. 
The \compOperationCopilot outperforms all the baselines by a large margin.
%; it does so through the use resolves these ambiguities by augmenting the same model with statistical profiles, achieving \textbf{100\% Semantic Validity} and \textbf{87\% Geometric Validity}.
Despite having access to dataset-level context, the Rule-Based system and  LLMs struggled with Geometric Validity, achieving 33\% and $\leq$74\%, respectively. This confirms that the dataset and column names are insufficient to reliably determine a variable's spatial representation. While all baselines exhibit low Geometric Validity (below 70\%), the Claude variants achieve high semantic validity (80\% and 94\%), demonstrating their sophisticated linguistic reasoning. 
%Semantic Validity via sophisticated linguistic reasoning, it still plateaus. 
The performance gap between the \compOperationCopilot and LLMs underscores the need for statistical profiling for robust spatial automation, enabling the system to move beyond the inherent ambiguity of municipal naming conventions to achieve "data-aware" reasoning.

\begin{table}[t]
  \centering
  \caption{Spatial operator selection accuracy. Geom.\ Val.\ denotes Geometric Validity (Mapping Operator); Sem.\ Val.\ denotes Semantic Validity (Aggregation Operator).}
  \label{tab:results}
  \resizebox{\columnwidth}{!}{%
  \begin{tabular}{lcc}
    \toprule
    \textbf{Condition}
      & \textbf{Geom.\ Val.\ (\%)}
      & \textbf{Sem.\ Val.\ (\%)} \\
    \midrule
    Rule-Based (Legacy GIS)                      & 33.0 &  28.0 \\
    Naive LLM (GPT-4o-Mini)                      &  7.0 &  51.0 \\
    Naive LLM (Claude 3.5 Sonnet)                & 52.0 &  80.0 \\
    Naive LLM (Claude 4.6 Opus)                  & 74.0 &  94.0 \\
    \textbf{\compOperationCopilot (Opus + Context)} & \textbf{87.0} & \textbf{100.0} \\
    \bottomrule
  \end{tabular}}
  \vspace{-.5cm}
\end{table}

\begin{figure}[!b]
  \centering
  \includegraphics[width=0.7\linewidth]{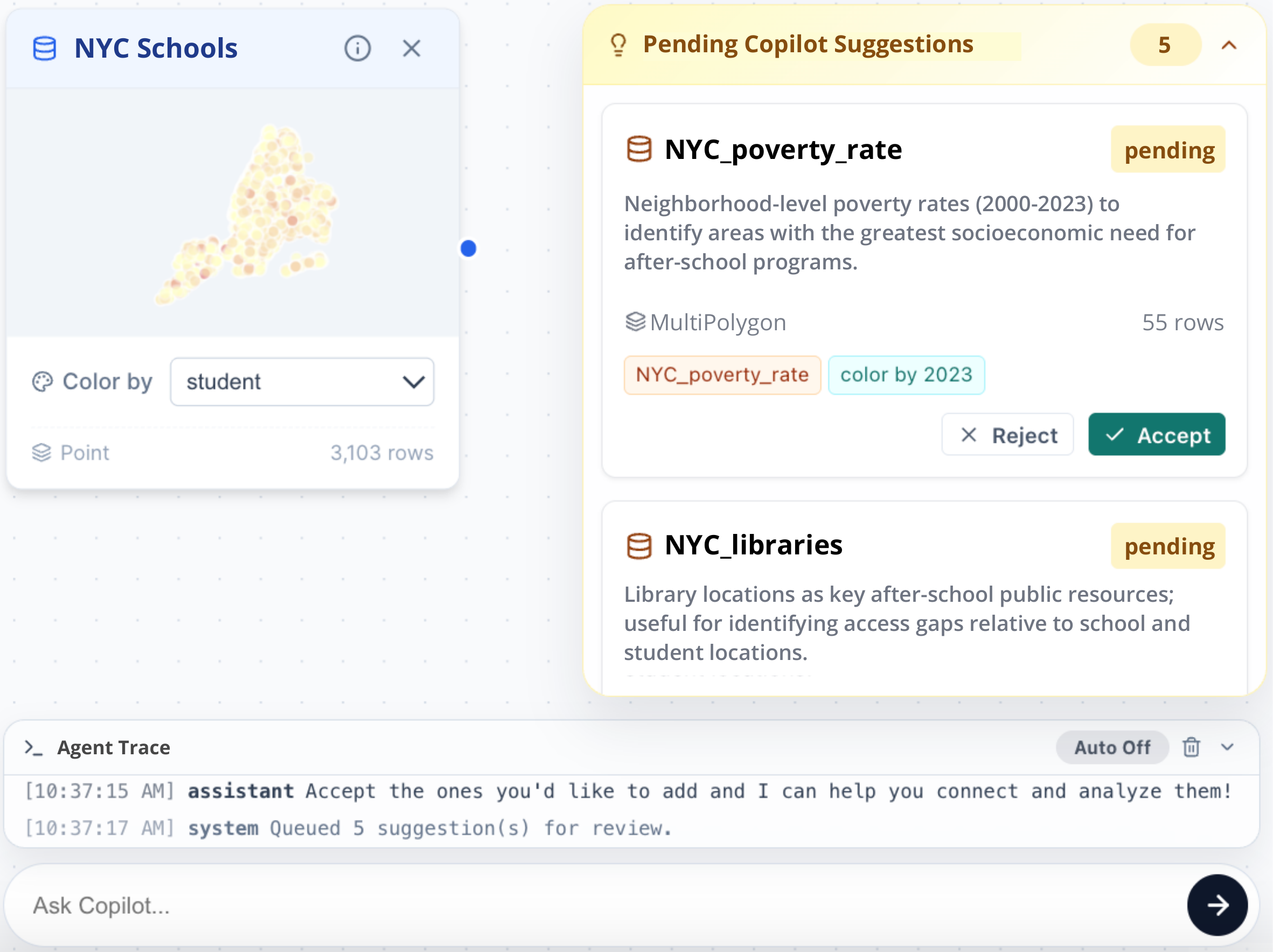}
  \vspace{-0.15cm}
  \caption{\compDiscoveryAgent in action. Based on the user's natural language prompt and the ``NYC Schools'' dataset already present on the canvas, the system suggests contextually relevant socio-economic and infrastructure datasets for the analysis.}
  \label{fig:discovery_canvas}
  \vspace{-0.5cm}
\end{figure}

\vspace{-0.15cm}

\subsection{\cmeasurement~Case Study 1: Digital Equity in the NYC Universal After-School Initiative}
\label{sec:cs1}

In 2025, the New York City Mayor’s Office launched an ambitious \emph{Universal After-School} initiative, announcing 115 new locations to bridge the city's educational divide \cite{nyc_after_school_2025}. While described as \emph{data-driven}, the initiative's spatial logic remained opaque. We aimed to determine if \name could \emph{replicate} these expert-level decisions using only raw open data, while producing a transparent, auditable digital twin of the decision-making process.

\paragraph{Intention-Driven Data Discovery.}
The workflow begins with the analyst formulating a high-level goal: identifying neighborhoods where after-school programs are most needed for educational and digital equity. The analyst starts with the most obvious foundational dataset, ``NYC Schools'' dataset (point geometries containing student enrolment capacity), dragging it onto the main canvas. To surface supporting variables, the analyst prompts \revision{the} \compDiscoveryAgent: \textit{``Find datasets related to socio-economic risk and access to public resources to study after-school program needs.''} Parsing the intent and the existing canvas context, the agent suggests median household income and poverty rates (Community District polygons) and public libraries and Wi-Fi hotspots (points) as shown in Figure~\ref{fig:discovery_canvas}.

% \begin{figure}[htbp]
%   \centering
%   \includegraphics[width=0.7\linewidth]{figs/casestudy1_discovery_edited.pdf}
%   \vspace{-0.15cm}
%   \caption{\compDiscoveryAgent in action. Based on the user's natural language prompt and the ``NYC Schools'' dataset already present on the canvas, the system suggests contextually relevant socio-economic and infrastructure datasets for the analysis.}
%   \label{fig:discovery_canvas}
%   \vspace{-0.4cm}
% \end{figure}

\paragraph{Data Heterogeneity and the Isolation Problem.}
The retrieved layers highlight a persistent challenge: they 
%are \emph{isolated} across 
\revision{have} incompatible coordinate systems, administrative boundaries (districts vs.\ neighborhoods), and statistical semantics. Naïve combination risks applying \emph{invalid operations}, e.g., summing a rate, thereby 
%silently 
corrupting downstream analysis. Figure~\ref{fig:teaser} (B3) illustrates how these choices, along with polarity adjustments, cascade into drastically different priority maps.

\begin{figure}[t]
  \centering
  \includegraphics[width=0.8\linewidth]{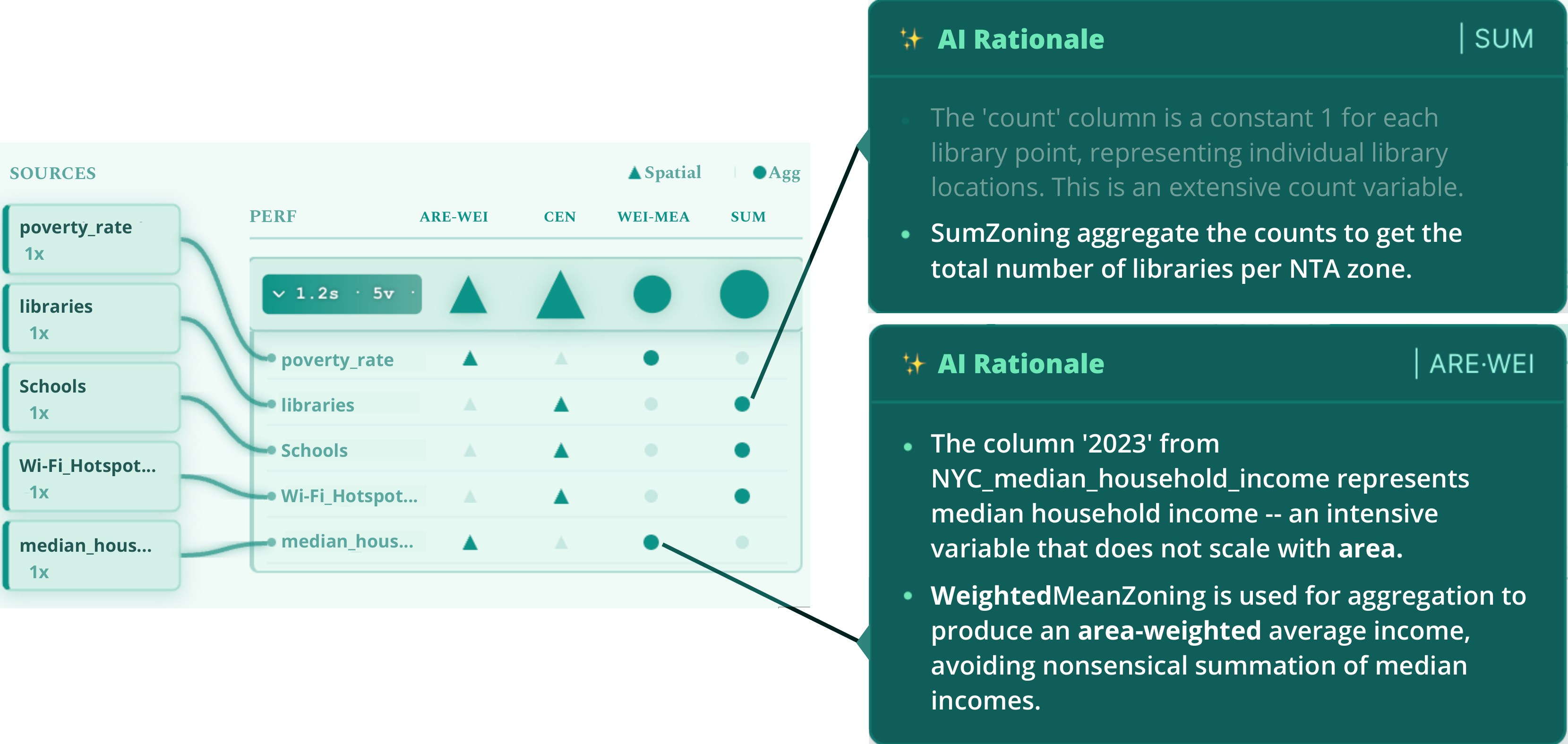}
\vspace{-0.15cm}
  \caption{Integration Provenance graph for the After-School case study.
           Each node encodes the semantic type of its variable and the operator
           recommended by the AI Copilot. The UI also exposes the AI rationale, 
           such as the logic behind selecting \texttt{sumZoning} for libraries 
           and \texttt{areaWeightedZoning} for household income.}
  \label{fig:topology}
  \vspace{-0.5cm}
\end{figure}

\paragraph{AI-Guided Semantic and Geometric Alignment.}
To resolve these conflicts, \revision{the} \compOperationCopilot first classifies each variable's semantic type
$C(v) \in \{\text{Extensive},\, \text{Intensive}\}$. 
It then selects the appropriate \textbf{spatial mapping} and \textbf{aggregation operators}: for the student-count (extensive point data), it applies a 
% \jf{spatial aggregation? over NTAs? } \scq{Yes, it is a spatial join to NTAs followed by aggregation. I revised the text accordingly.}
\emph{spatial join} to \revision{NYC Neighborhood Tabulation Areas (NTAs), followed by aggregation (sum) within each NTA}; for poverty rates (intensive district polygons), it applies \revision{\emph{Areal Weighted Interpolation} (AWI)} to re-project values onto \revision{the same spatial units} without distorting magnitudes. Unlike traditional GIS workflows, the resulting \emph{Integration Provenance \revision{Graph}} (Figure~\ref{fig:topology}) makes every transformation step and AI rationale, such as the choice of \texttt{areaWeightedZoning} for income, fully \revision{transparent} and traceable.

\begin{figure}[!b]
  \centering
  \begin{minipage}{0.58\linewidth}
    \centering
    \includegraphics[width=\linewidth]{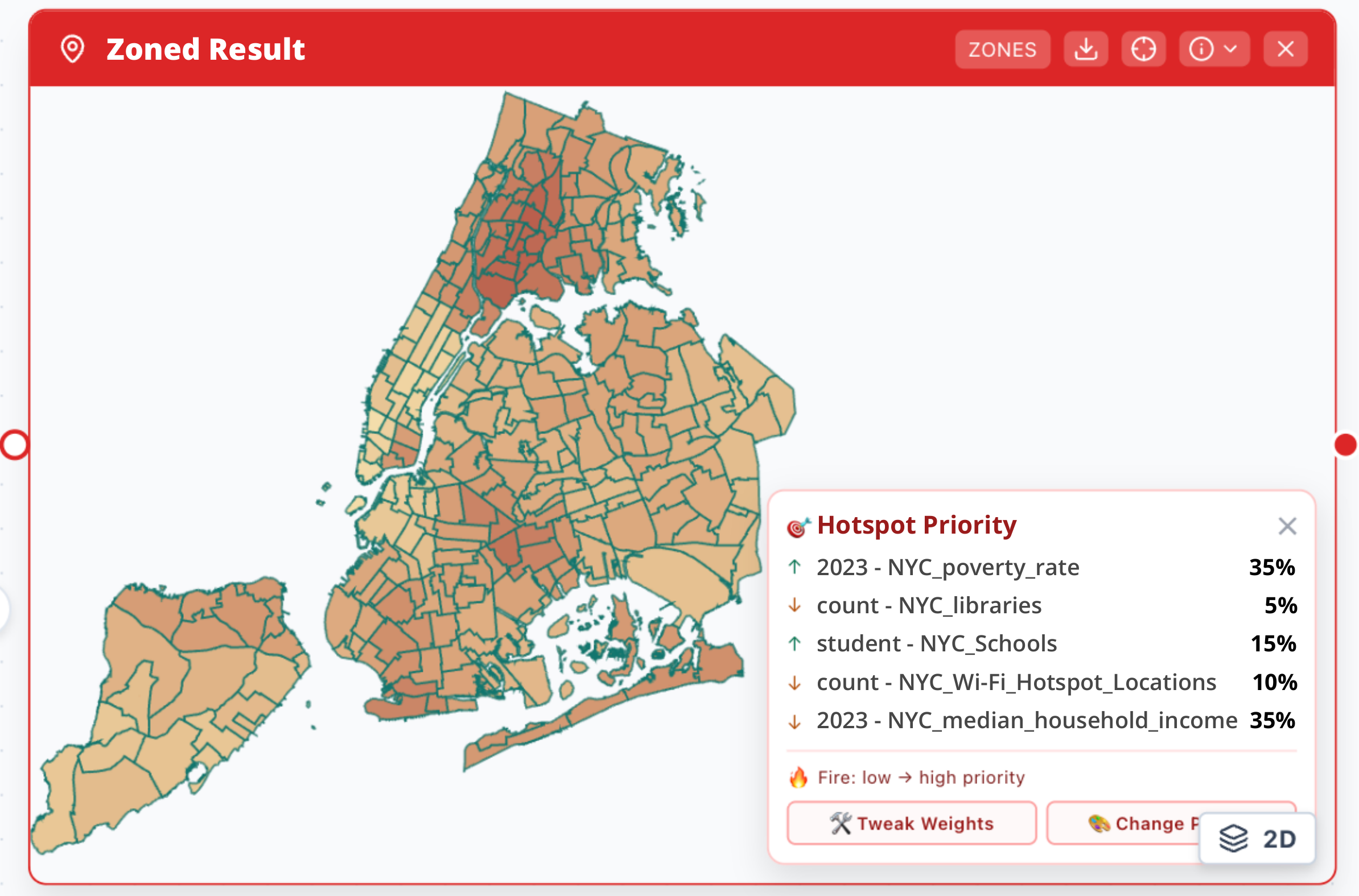}
  \end{minipage}
  \hfill
  \begin{minipage}{0.36\linewidth}
    \centering
    \includegraphics[width=\linewidth]{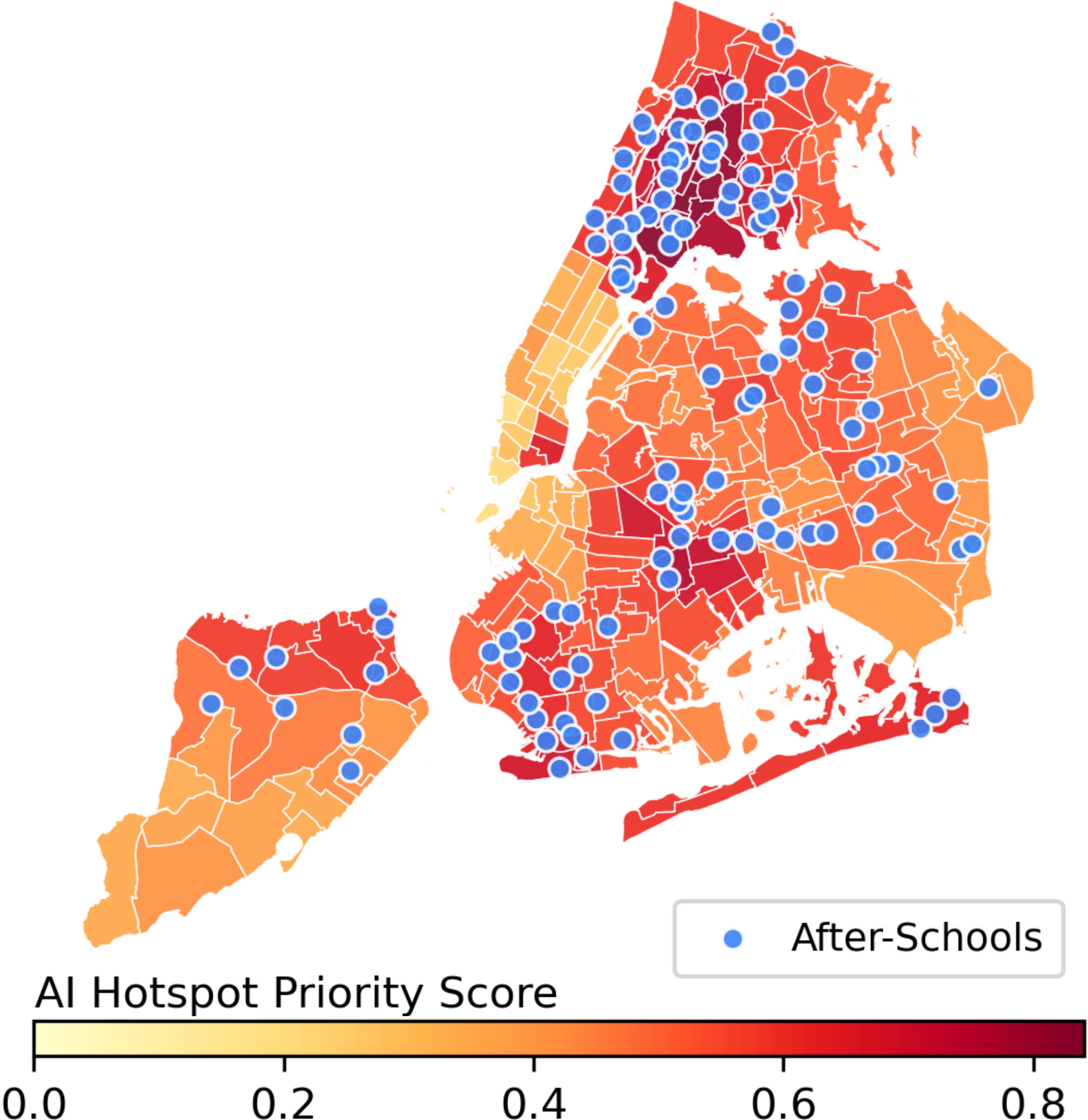}
  \end{minipage}
\vspace{-0.15cm}
  \caption{Spatial distribution of the Multivariate Hotspot Index and ground-truth validation. 
           \textbf{Left:} The composite index $P$ \hide{(Equation~\ref{eq:hotspot})} rendered as a choropleth over NYC \revision{NTAs}, where \emph{deep red} districts score highest. The inset legend (bottom-right) details the parameter configuration, including inverted polarity for public Wi-Fi, libraries, and median household income, with user-adjustable weights.
           \textbf{Right:} AI Hotspot Priority Zones overlaid with the 114 geocoded Universal After-School locations used for validation.}
  \label{fig:hotspot_combined}
  \vspace{-0.5cm}
\end{figure}

\paragraph{Hotspot Synthesis and Priority Visualization.}
Once harmonized, the framework computes a \textit{Multivariate Priority Map} ($P$)
\hide{at the \textsf{Result Map Node}}. The LLM infers the \emph{directional polarity} required to surface the \emph{opportunity gap}, the co-occurrence of high demand, high vulnerability, and limited services: \revision{$
P = w_{1}\tilde{V}_{\text{students}} +
w_{2}\tilde{V}_{\text{poverty}} +
w_{3}(1-\tilde{V}_{\text{income}}) +
w_{4}(1-\tilde{V}_{\text{wifi}}) +
w_{5}(1-\tilde{V}_{\text{library}})
$,}
% \begin{small}
% \begin{equation}
% P = w_{1}\tilde{V}_{\text{students}} + w_{2}\tilde{V}_{\text{poverty}} + w_{3}(1 - \tilde{V}_{\text{income}}) + w_{4}(1 - \tilde{V}_{\text{wifi}}) + w_{5}(1 - \tilde{V}_{\text{library}})
% \label{eq:hotspot}
% \end{equation}
% \end{small}
% \noindent 
where $\tilde{V}$ denotes min-max normalized values. By \emph{inverting} income and infrastructure, the system highlights \emph{underserved areas}, areas where elevated demand and concentrated poverty intersect with a deficiency of existing services (Figure~\ref{fig:hotspot_combined}, Left). While the LLM proposes initial weights $w_{i}$ based on the policy objective, the analyst can manually refine them \hide{via interactive sliders} to explore alternative scenarios.

\paragraph{Validation Against Ground Truth.}
To validate the framework, we geocoded the official addresses of the 115 Universal After-School sites announced by the Mayor's Office\revision{, successfully locating 114 sites as ground truth} (Figure~\ref{fig:hotspot_combined}, Right). \revision{Using percentile thresholds (Table~\ref{tab:validation}), \name captures \textbf{77.2\%} of sites within the top 50\% of priority zones (lift = 1.54).
%; the top 30\% captures 56.1\% (lift = 1.87), confirming strong enrichment of official intervention sites.
As selectivity increases, the top 30\% of zones capture 56.1\% of sites (lift = 1.87), 
confirming that higher-ranked zones are significantly enriched with actual intervention sites.
These results show that an AI-assisted pipeline can closely reproduce municipal site-selection decisions. Unlike black-box prediction, \name records every weight, operator, and boundary transformation in the topology graph, providing a transparent chain of evidence that supports verifiable public policy decisions.}

% \paragraph{Validation Against Ground Truth.} To validate the framework, we geocoded the official addresses of the 115 Universal After-School sites announced by the Mayor's Office. We successfully isolated exact locations for 114 sites, which serve as our ground-truth analytical dataset (Figure~\ref{fig:hotspot_combined}, Right). Evaluating priority quality via percentile thresholds (Table~\ref{tab:validation}), \name captures \textbf{77.2\%} of official locations within the top 50\% of priority zones (lift = 1.54). As selectivity increases, the top 30\% of zones capture 56.1\% of sites (lift = 1.87), confirming that higher-ranked zones are significantly enriched with actual intervention sites. These results demonstrate that an AI-assisted pipeline can replicate the complex municipal decisions with high fidelity. \revision{Furthermore,}  \name moves beyond "black-box" prediction by providing a documented \emph{chain of evidence}: every weight, operator, and boundary transformation is recorded in the topology graph. This level of \revision{transparency} allows urban analysts to justify and defend site-selection results, transforming automated spatial discovery into a verifiable instrument for public policy.

\begin{table}[h]
  \centering
  \small % Slightly scales down the font to better fit the column
  \setlength{\tabcolsep}{4pt} % Squeezes the whitespace between columns
  % \caption{Quantitative validation of hotspot quality across percentile thresholds.}
  \caption{Hotspot quality validation across percentile thresholds.}
  \vspace{-0.2cm}
  \label{tab:validation}
  \begin{tabular}{@{} l c c c c c c @{}} % The @{} removes the invisible padding on the far left and right
    \toprule
    \textbf{Quantile} & \textbf{Thresh.} & \textbf{Coverage} & \textbf{$N_{\text{high}}$} & \textbf{$N_{\text{total}}$} & \textbf{Baseline} & \textbf{Lift} \\
    \midrule
    0.5 & 0.530 & 77.2\% & 88 & 114 & 0.5 & 1.54 \\
    0.6 & 0.557 & 69.3\% & 79 & 114 & 0.4 & 1.73 \\
    0.7 & 0.584 & 56.1\% & 64 & 114 & 0.3 & 1.87 \\
    \bottomrule
  \end{tabular}
  \vspace{-.4cm}
\end{table}

\subsection{\cspatialunit~Case Study 2: Uncovering "Opportunity Bargains" via Interactive Scenario Testing}
\label{sec:cs2}
% ------------------------------------------------------------------
\begin{figure*}[htbp]
  \centering
  \includegraphics[width=.9\textwidth]{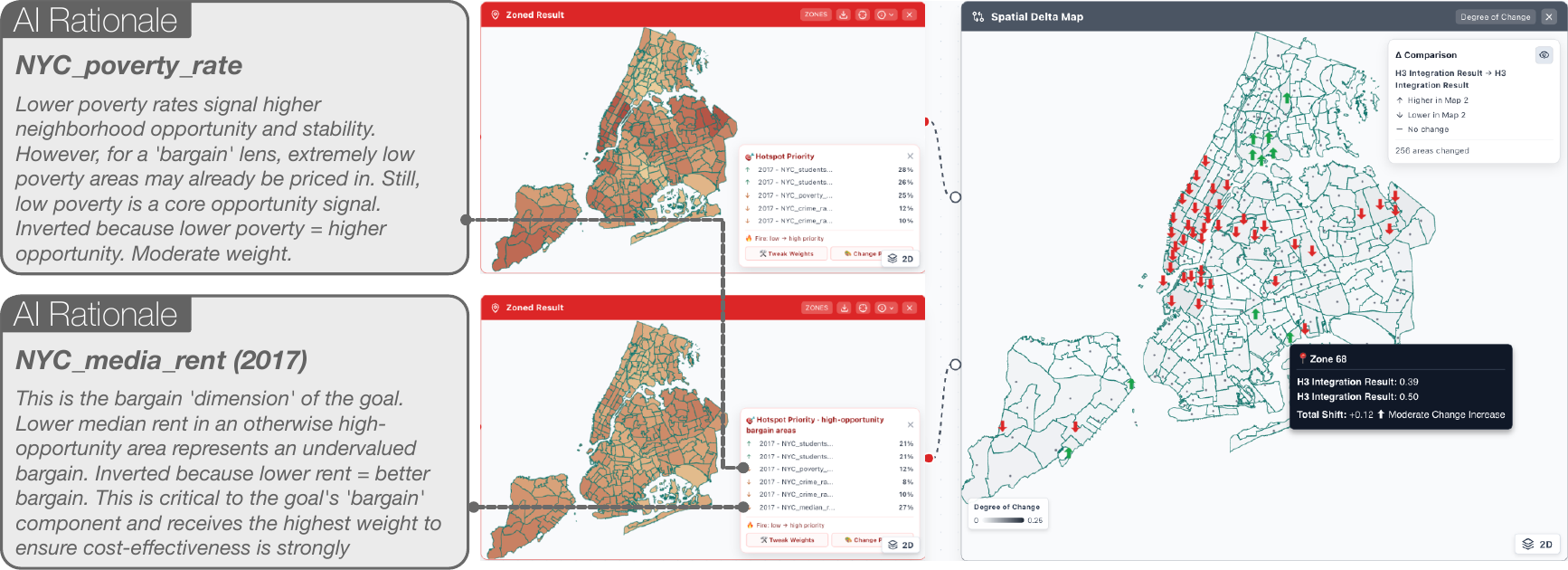}
  \caption{Exploratory Scenario Testing via the \emph{Spatial Delta Map}. \textbf{Top-Middle:} The baseline SVP index map, highlighting high-opportunity areas regardless of housing cost. \textbf{Bottom-Middle:} The modified index map (SVP + Rent), incorporating dynamically harmonized Sub-Borough rent data. The annotated tooltips (\textbf{Left}) provide examples of the AI Copilot's semantic reasoning, demonstrating how it automatically applies inverted polarity (shown here for poverty and median rent) to maximize opportunity. \textbf{Right:} The resulting Spatial Delta Map. Areas marked by \emph{red downward arrows} indicate neighbourhoods that dropped in priority due to prohibitive rents, while \emph{green upward arrows} highlight newly surfaced ``Opportunity Bargains.''}
  \label{fig:spatial_delta}
  \vspace{-0.6cm}
\end{figure*}

% \begin{figure*}[htbp]
%   \centering
%   \includegraphics[width=0.95\textwidth]{figs/casestudy2_spatial_delta_v2.png}
%   \caption{Exploratory Scenario Testing via the \emph{Spatial Delta Map}. \textbf{Top-Middle:} The baseline SVP index map, highlighting high-opportunity areas regardless of housing cost. \textbf{Bottom-Middle:} The modified index map (SVP + Rent), incorporating dynamically harmonized Sub-Borough rent data. The annotated tooltips (left) provide examples of the AI Copilot's semantic reasoning, demonstrating how it automatically applies inverted polarity (shown here for poverty and median rent) to maximize opportunity. \textbf{Right:} The resulting Spatial Delta Map. Areas marked by \emph{red downward arrows} indicate neighbourhoods that dropped in priority due to prohibitive rents, while \emph{green upward arrows} highlight newly surfaced ``Opportunity Bargains.''}
%   \label{fig:spatial_delta}
% \end{figure*}
\paragraph{Motivation: Overcoming Rigid Pipeline Constraints.}
Urban planners \revision{often use} composite indices to allocate resources or guide housing policy. \revision{One} example is the School-Violence-Poverty (SVP) index~\cite{kelly2023planning}, which identifies neighborhoods maximizing economic mobility for children. Planners use these measures to allocate subsidies and \hide{direct} affordable housing, while families use them 
%to find neighborhoods with strong opportunities at reasonable cost. 
\revision{to identify high-opportunity neighborhoods.}
Traditionally, "opportunity bargains" are \revision{found} by comparing SVP scores against housing costs using post-hoc model residuals.

% \paragraph{Motivation: The "Static Model" Trap.} Urban planners frequently rely on composite indices to allocate resources or guide housing policy. A prominent contemporary example is the School-Violence-Poverty (SVP) index proposed by Kelly and Ellen~\cite{kelly2023planning}, which identifies neighborhoods that maximize economic mobility for children. The SVP index and subsequent opportunity bargain analysis are of direct interest to two key groups: (1) planners, who can use these measures to target public subsidy dollars and guide affordable housing development toward high-opportunity areas, and (2) families with children, who can use them to identify neighborhoods that offer strong life outcomes at attainable costs. In prior work~\cite{kelly2023planning}, opportunity bargains are identified by comparing SVP scores against expected housing costs, using model residuals to surface neighborhoods where rents are lower than predicted.

However, conventional GIS software 
%often traps analysts in \emph{rigid, fixed workflows}: Domain-specific models like SVP are predefined into linear scripts. If a policymaker wishes to perform a counterfactual analysis, such as incorporating affordability into the SVP model, analysts must manually acquire new data, resolve spatial geometry mismatches (e.g., Zip Codes versus Census Tracts), and 
% \jf{write? update?} \scq{'update' would be more appropriate here. For more precision, we can say 'modify and rerun,' since analysts typically adjust existing scripts rather than writing them from scratch.}
% \revision{update} transformation scripts. This rigid pipeline prevents analysts from reformulating the model itself during analysis, limiting their ability to explore alternative definitions of “opportunity” under evolving policy constraints. In this case study, we demonstrate how our framework replaces static model execution with interactive AI-guided spatial scenario testing.
\revision{are largely fixed. Incorporating new factors, such as affordability, requires acquiring additional datasets, resolving spatial mismatches (e.g., Zip Codes versus Census Tracts), and updating transformation scripts. This prevents analysts from interactively reformulating the model during analysis to explore alternative definitions of opportunity. This case study demonstrates how our framework instead supports AI-guided spatial scenario testing.}

\paragraph{Dynamic Construction of the Baseline Index.}
Instead of relying on a pre-computed shapefile, the user constructs the baseline SVP index interactively within the framework. The user imports five datasets \revision{obtained} from the Furman Center~\cite{furmancenter2024}:
% \jf{data collection?} \scq{I revised 'datasets sourced from' to 'datasets obtained from' for clarity. Also, I clarified that geometries are not included in the original datasets and are attached by merging with an official shapefile.}
elementary student performance (Math and English Language Arts) and crime rates (violent and property crimes), both aggregated at the Community District level, alongside poverty rates aggregated at the Sub-Borough Area level. 
% \scq{The data are then merged with an official shapefile to attach geometries.} 
These datasets
%represent the school, violence, and poverty 
include the metrics needed to \revision{compute} 
% \jf{compute?} \scq{I revised to “compute” here because we are actually constructing the SVP index from the datasets. I still use “approximating” later when referring to prior work since we keep the same structure and polarity, but replace the original Z-score standardization with min–max normalization. This gives us a clean 0–1 scale for visualization while keeping the overall idea of the SVP index. In future versions, we could switch to Z-scores to better handle outliers, but then we would need to convert them into percentile ranks (0–100) for visualization.} 
the core SVP concept. 
% Community District and Sub-Borough Area shapefiles from NYC Department of City Planning (https://www.nyc.gov/content/planning/pages/resources/datasets/community-districts) and NYU CUSP (https://geo.nyu.edu/catalog/sde-columbia-cul_2004_nyc_subborough)

Although all inputs are polygon layers, they represent fundamentally \revision{different} administrative boundaries. To support both policy-level targeting and household-level exploration, the user sets the target spatial unit to \revision{NTAs}, which provide a neighborhood-scale resolution suitable for both planning and decision-making contexts. The AI Copilot recognizes these mismatched geometries and automatically applies an \revision{AWI} operator to harmonize both the Community District and Sub-Borough Area polygons into the NTA spatial resolution. The baseline \texttt{Result Map Node} then computes the SVP hotspot priority map (Figure~\ref{fig:spatial_delta}, Top-Middle), approximating the spatial patterns reported in prior work~\cite{kelly2023planning} by highlighting traditionally high-opportunity neighborhoods.

\paragraph{The ``Injection'': Flexible Spatial Harmonization.}
To transform the analysis from a static observation of opportunity into an actionable decision-support tool for both planners and families, the user dynamically injects a sixth dataset into the live topology \revision{visualization}%\jf{visualization} \scq{updated}
: Median Rent, which is also aggregated at the Sub-Borough Area level. 

In a traditional GIS environment, integrating this new variable into an NTA-level index would 
%halt the analysis, requiring 
\revision{require} the user to manually script another round of spatial joins and areal interpolation. In our framework, the AI Copilot identifies the geometric conflict and automatically inserts the appropriate AWI operator to harmonize the rent data \revision{with} the unified NTA map. Furthermore, the Copilot infers the analytical intent (identifying affordable, high-opportunity areas) and applies an \emph{inverted directional polarity} to the rent variable, ensuring that lower rents increase the composite hotspot score (Figure~\ref{fig:spatial_delta}, Bottom-Middle). 
While this formulation departs from the original residual-based definition of opportunity bargains, it enables a complementary perspective in which affordability is treated as an intrinsic component of opportunity rather than a post hoc adjustment.

\paragraph{Exploratory Analysis via the Spatial Delta Map.}
Because our framework supports interactive model construction, the user can immediately visualize the impact of the rent \revision{layer} using a \emph{Spatial Delta Map} (Figure~\ref{fig:spatial_delta}, Right). 
%
%The delta visualization encodes the marginal effect of affordability on opportunity, revealing how the inclusion of rent reshapes the spatial distribution of priority areas. Highly gentrified neighborhoods that scored well on the baseline SVP index exhibit a sharp negative delta (\emph{red down arrows}), indicating areas where high opportunity is offset by prohibitive housing costs, critical signals for planners evaluating where subsidies may be necessary to preserve access. Conversely, hidden ``opportunity bargains'' emerge as positive deltas (\emph{green up arrows}), areas where strong SVP characteristics are amplified by relatively lower rents, making them attractive both for targeted development strategies and for families seeking affordable access to opportunity.
\revision{The map encodes the marginal effect of affordability on opportunity. High-opportunity but expensive neighborhoods appear as negative deltas (\emph{red down arrows}), indicating where housing costs diminish opportunity and subsidies may be needed. Conversely, positive deltas (\emph{green up arrows}) reveal hidden ``opportunity bargains,'' where strong SVP characteristics coincide with relatively affordable housing.}

This workflow demonstrates a fundamental shift in spatial analytics. By automating the mechanics of operator selection and polygon-to-polygon harmonization, the framework elevates the analyst from executing rigid, one-off geoprocessing pipelines to engaging in rapid, domain-specific hypothesis testing and model reformulation across both policy and individual decision-making contexts.

%%%%%%%%%%%%%%%%%%%%%%%%%%%%%%%%%%%
%%%%% Expert Interviews
%%%%%%%%%%%%%%%%%%%%%%%%%%%%%%%%%%%
\subsection{Domain Expert \revision{Interview}}
\label{sec:expert_feedback}

To assess the practical utility of \name and evaluate how its AI-assisted pipeline integrates into real-world analytical workflows, we conducted \hide{in-depth, }semi-structured expert evaluation sessions. \revision{Per design-study methodology~\cite{sedlmair2012}, these sessions serve as a strictly summative validation of the completed system, rather than formative co-design.}

\paragraph{Participants and Protocol.}
We recruited four domain experts in urban data:
% experts with extensive experience in urban data analytics:
an MIT CS/City Planning researcher (E1); the Assistant Director of Research at a university transportation \revision{center} (E2); an urban data software engineer (E3); and an urban researcher from the NYU Center for Urban Science (E4).
%with a background in applied AI and urban mobility (E4). 
% Each session lasted 60 to 100 minutes. Following a system introduction (15 min), experts were asked to complete a think-aloud spatial integration task (based on Case Study 1) (35 min), and a semi-structured interview focusing on usability, trust, and workflow integration (10 min).
Sessions lasted 60--100 minutes and included a system introduction (15 min), a think-aloud spatial integration task based on Case Study 1 (35 min), and a semi-structured interview evaluating usability, trust, and workflow integration (10 min). %\revision{All participants provided informed consent prior to participating in the study.}

\revision{Domain experts provided professional feedback on urban data analysis workflows and the relevance of UrbanTrace's capabilities. Because the study gathered professional domain expertise rather than studying the experts themselves, IRB review was not required under NYU guidance. All participants provided informed consent prior to participation.}

\paragraph{Workflow Acceleration.}
All experts \revision{emphasized} that \name accelerates the early stages of spatial analysis. E1, whose current workflow relies on manually coding Python and GeoPandas scripts, noted that she ``can't imagine how much time'' the system would save, describing the visual transition from data discovery to integration as ``straightforward'' and ``compelling.'' E2 echoed this sentiment, likening the drag-and-drop canvas paradigm to familiar business intelligence tools (e.g., Tableau), which she noted lowers the barrier to entry for urban planners who lack advanced programming skills. E4 first used keyword search but quickly switched to the \compDiscoveryAgent, observing that natural-language retrieval streamlined exploration.

\paragraph{Intent-Driven Discovery and Data Exploration.}
Before executing integrations, analysts must \revision{locate} and verify \revision{the} relevant data. The experts highlighted the \revision{\compDiscoveryAgent} as a major improvement over rigid open data portals. E2 emphasized that real-world dataset column names are often obscure; \revision{E2} and E4 both found the system's auto-generated dataset summaries and sample records embedded directly in the \textit{Dataset Nodes} crucial for interpreting \revision{the dataset} structure %\jf{the dataset structure} \scq{updated}
\emph{before} selection. 
% To further lower the cold-start barrier, E3 suggested adding example prompts to help novice users initiate the discovery agent.
Once \revision{the} datasets were retrieved, the visual exploration features proved \revision{to be} effective. E1 and E4 noted that synchronized map navigation \revision{enabled} them to seamlessly verify local data distributions across multiple layers \revision{simultaneously}, ensuring contextual alignment before invoking the AI Copilot for formal integration.

\paragraph{Integration Strategy Generation and Mitigating the ``Black Box.''} 
While experts (E2, E3) explicitly warned against ``black box'' systems for spatial integration, they validated that \name's architecture successfully mitigates these concerns. 
%From an engineering perspective, 
E3 initially feared computational and hallucination risks but acknowledged the \compAtlas's metadata injection as an effective optimization that grounds the LLM.
% Experts expressed skepticism toward LLM-generated spatial integration strategies, warning against “black box” systems (E2, E3). 
%A recurring theme across all sessions was the inherent skepticism domain experts hold toward LLMs autonomously generating spatial integration strategies. Both E2 and E3 explicitly warned against ``black box'' systems.
% The experts validated that \name's architecture successfully mitigates these concerns. E3, approaching the system from an engineering perspective, initially expressed concern about the computational cost and hallucination risk, but upon reviewing the \compAtlas's metadata-injection architecture, he acknowledged it as an effective optimization that grounds the LLM. 
On the frontend, the \emph{Integration Provenance} \revision{\emph{Graph}} served as the primary anchor for user trust, with E4 noting that exposing the LLM rationales and making the integration logic visible supports user engagement and trust.
% the \emph{Integration Provenance} graph served as the primary anchor for user trust. E4 responded positively to the provenance panel exposing the LLM rationales, noting that making the integration logic visible supports user engagement and trust. 
E1 highlighted that navigating mismatched Coordinate Reference Systems (CRS) and partial geometric overlaps is a ``big problem'' in practice. While she found the provenance graph ``made [her] feel very controllable,'' \revision{she still wanted to verify} 
%\jf{this is not a proof, maybe "she still wanted to verify the computation, and so did E4"} \scq{I rephrase this to avoid the term “proof,” since E1 and E4 were not asking for a formal mathematical proof, but rather wanted to verify how the spatial mapping was performed and understand the intermediate process (grid construction).}
\revision{the underlying computations and spatial transformations, as did E4}. When shown the system's ability to expose the intermediate grid-based allocations, E1 responded emphatically: ``That is exactly what I want.'' This highlights a critical HCI finding for spatial AI: structural provenance (the node graph) must be paired with geometric provenance (inspectable intermediate grid\revision{-based visualization}) to secure expert trust. 
% \jf{it is not clear what the intermediate grid is -- need to explain } \scq{I clarify this by explicitly referring to the inspectable intermediate grid-based visualization that represents the grid-based allocations (as described in Section 3.6). \revision{The intermediate grid-based visualization} is not shown in the paper due to space constraints.}

\paragraph{Interpretable Spatial Synthesis.}
The experts actively engaged with the \emph{Multivariate Priority Map} and the \emph{Spatial Delta Map}, testing counterfactual scenarios by adjusting variable weights. E1 praised the LLM's automated inference of directional polarity (e.g., inverting infrastructure variables to highlight gaps), describing it as ``very straightforward and easy to understand'' for policy planning. E2 initially raised concerns about mathematically combining skewed distributions (e.g., raw student counts vs. library counts), but was reassured by the system's explicit display of its backend normalization logic and the LLM rationale tooltips. E3 described the end-to-end synthesis as ``crazy cool,'' noting that deriving multi-factor insights from raw data is %historically 
non-trivial even for experienced programmers. 
E4 found the \textit{Delta Map}’s directional arrows effective for tracking hotspot shifts, calling the dynamic weight adjustments ``really helpful'' for what-if exploration.
% E4 showed a pronounced positive reaction to the \textit{Delta Map}’s directional arrows, noting the visual encoding effectively communicated how new hotspots emerge or disappear as parameters change. E4 characterized the ability to dynamically tweak weights as ``really helpful,'' highlighting the system's support for intuitive what-if exploration.

\paragraph{Expert Perspectives on \name Extensions.}
Suggested extensions include time-series operators for temporal dynamics (E3, E4), LLM-assisted geocoding \revision{of locations from legacy administrative records (e.g., datasets with free-text addresses)} (E2), 
% \jf{what do you mean by geocoding noisy legacy data? concrete example?} \scq{I clarify this by referring to LLM-assisted geocoding of legacy administrative records (e.g., datasets with free-text addresses), where location information is not explicitly structured as coordinates and must be extracted and converted into geometry.}
and \revision{code-based preprocessing within the tool prior to integration} for advanced research (E4). 
% \jf{what exactly does this user want? to input the formulas?} \scq{By “direct code manipulation,” E4 refers to using Python to preprocess and modify datasets prior to ingestion into the system’s integration pipeline. For example, she would like to perform data preprocessing at the Dataset Node before integrating it with other datasets. I revised this.}
%\jf{Too many 'crucially' in the paper, e.g., here it is not necessary -- could just start at "To ensure...."} \scq{updated. Thanks!}
\revision{To} ensure policy defensibility and provide an auditable paper trail of the ``reasoning process'' (E1), experts (E1, E3, E4) recommended exporting the \textit{Integration Provenance} alongside LLM-generated narrative summaries for seamless embedding into municipal reports.
\vspace{-0.15cm}

\section{\revision{Discussion and Limitations}}
\label{sec:discussion}

% \jf{what do you mean by automated error-blocking?}
\paragraph{\revision{Hallucination Mitigation and Dataset Quality.}}
\revision{LLM-driven spatial workflows risk introducing silent errors via hallucination and bias. \name mitigates this structurally: ingestion quality is partially verified through rule-based geometry checks (though source accuracy remains the portal's responsibility), while \compOperationCopilot is confined to a strict operator space and \compAtlas grounds every LLM call in observed statistics rather than free-form generation. Rather than relying on automated error-blocking, we enforce human-in-the-loop transparency: the \emph{Provenance Graph} exposes all structural inferences, tooltips display AI-generated rationales, and the interface allows analysts to override incorrect operator choices, weights, or polarities, transforming potential silent failures into inspectable choices.}

\paragraph{\revision{MAUP and Spatial Aggregation.}}
\revision{UrbanTrace does not eliminate MAUP; no single spatial partition can. Rather, the system makes boundary sensitivity \textit{explorable}: the grid resolution is
user-configurable, and the \emph{Spatial Delta Map} exposes how synthesized priorities shift under alternative configurations. As discussed in Section~\ref{sec:formal_model}, the common grid serves as a geometry-agnostic intermediate layer that reduces misalignment errors between incompatible administrative boundaries, while MAUP-related effects remain dependent on the analyst's chosen resolution and reporting units.}

\paragraph{\revision{Limitations.}}
\revision{
1)~While \compOperationCopilot achieved 100\% semantic validity, geometric validity reached 87\%. %The remaining 13\% reflects localized boundary-alignment challenges in a subset of polygon-to-linear tasks rather than systematic reasoning failures, occasionally requiring manual intervention.
The remaining 13\% reflects localized boundary-alignment challenges 
during spatial joins across complex or unaligned geographic scales, such as linear network features with polygon zones, %when integrating linear network features with polygon zones,
rather than systematic reasoning failures, occasionally requiring manual intervention.
2)~Offline profiling limits adaptability to real-time\hide{or unstructured data}, requiring future dynamic profiling. %\jf{why mention unstructured data? this sounds confusing here without additional context; and we are focusing on tabular data}
3)~\compDiscoveryAgent currently processes entire catalogs, incurring linear token costs; future work will integrate retrieval methods to pre-filter datasets.
4)~The \emph{Integration Provenance Graph} introduces a learning curve over conventional chat interfaces; however, expert feedback (Sec.~\ref{sec:expert_feedback}) suggests this trade-off is justified by the transparency and trust provided through explicit visual provenance.}

\vspace{-0.15cm}

\section{Conclusion}
\label{sec:conclusion}
Integrating heterogeneous spatial data has traditionally required urban analysts to manually bridge the gap between high-level policy goals and rigid spatial operations. In this paper, we present \name, a visual analytics system that fills this gap by coupling LLMs with structured
data profiling, transforming opaque spatial data wrangling into a transparent, intention-driven collaborative workflow \revision{that enables intent-driven data discovery and semantic-aware harmonization.}

\paragraph{Future Work.}
Building on these findings, future research will focus on expanding the system's spatial reasoning capabilities. We plan to integrate more sophisticated harmonization techniques, such as allowing \compOperationCopilot to recommend auxiliary-variable dasymetric interpolation, which provides higher fidelity than standard areal weighting. We aim to extend \compCanvas to support spatio-temporal analysis, enabling users to visualize how spatial priorities evolve
over time. Finally, we plan \revision{to evaluate \name across diverse geographic regions and} scale the \compAtlas and \compDiscoveryAgent \revision{to support multiple heterogeneous municipal data lakes, enabling large-scale cross-city comparative analytics.}

\vspace{-.15cm}
\section*{Supplemental Material}{
Details regarding specifications and prompt templates for UrbanTrace’s three LLM agents, 28 benchmark scenarios, ablation studies, and complete source code are available in our public repository release \cite{urbantrace_supplementary}.
}

\acknowledgments{%
The authors thank the reviewers for the constructive comments and the domain experts that evaluated the system. This work was supported in part by the National Science Foundation grant OAC-2411221. Silva was partially supported by the DARPA expMath program (under cooperative agreement HR0011262E029).
Any opinions, findings, and conclusions or recommendations expressed in this material are those of the authors and do not necessarily reflect the views of  NSF or DARPA.}

\bibliographystyle{abbrv-doi-hyperref}

\bibliography{references}

\end{document}